\documentclass[reprint,showpacs,preprintnumbers,amsmath,amssymb, prl]{revtex4-1}
\usepackage{graphicx,wasysym}
\usepackage{dcolumn}

\usepackage{bm}
\begin{document}

\preprint{1}

\title{Nonlinear transport coefficients from large deviation functions}

\author{Chloe Ya Gao}
\author{David T. Limmer} 
\affiliation{Department of Chemistry, University of California, Berkeley, CA 94609}
\affiliation{Kavli Energy NanoScience Institute, Berkeley, CA 94609}
\affiliation{Materials Science Division, Lawrence Berkeley National Laboratory, Berkeley, CA 94609}
 \email{dlimmer@berkeley.edu}


\begin{abstract}
Nonlinear response occurs naturally when a strong perturbation takes a system far from equilibrium. Despite of its omnipresence in nanoscale systems, it is difficult to predict in a general and efficient way. Here
we introduce a way to compute arbitrarily high order transport coefficients of stochastic systems, using the framework of large deviation theory. Leveraging time reversibility in the microscopic dynamics, we relate nonlinear response to equilibrium multi-time correlation functions among both time reversal symmetric and asymmetric observables, which can be evaluated from derivatives of large deviation functions. This connection establishes a thermodynamic-like relation for nonequilibrium response and provides a practical route to its evaluation, as large deviation functions are amenable to importance sampling. We demonstrate the generality and efficiency of this method in predicting transport coefficients in single particle systems and an interacting system exhibiting thermal rectification.

\end{abstract}



\maketitle


Transport processes are often described by phenomenological laws, such as Fourier's law for heat conduction, Fick's law for diffusion, and Ohm's law for electrical conduction. These equations propose a relationship between the thermodynamic force, or affinity, that drives the system out of equilibrium and the resultant current, where the strength of the response is described by a material specific transport coefficient. While connections between these coefficients and the underlying microscopic degrees of freedom are well established when forces are small, such connections are much less clear when forces are large and responses are nonlinear. The purpose of this paper is to develop a general scheme for the calculation of arbitrarily high order transport coefficients in stochastic systems from microscopic equilibrium fluctuations, and illustrate its viability by computing nonlinear transport coefficients from the derivatives of large deviation functions.

The notion of relating transport coefficients to spontaneous fluctuations in equilibrium dates back to Onsager's regression hypothesis, which states that the relaxation of a system from a given nonequilibrium state prepared by an external perturbation, obeys the same law as the regression of the same state produced by spontaneous thermal fluctuations \cite{onsager1931reciprocal}. A further connection between the two was established later by deriving
the path probability of a given succession of states in a system where the macroscopic variables are Gaussian distributed \cite{onsager1953fluctuations}. Along the same line, Kubo constructed a general statistical-mechanical theory for irreversible processes \cite{kubo1957statistical}. The well known Green-Kubo formulas express a linear response coefficient with the integral of the current autocorrelation function over time, and have become a standard method to compute linear transport coefficients \cite{green1954markoff,levesque1973computer,schelling2002comparison}. Onsager and Kubo's work are the cornerstones of linear response theory, which applies to irreversible processes close to equilibrium.

As the external perturbation grows stronger, 
nonlinear transport behaviors naturally arise, such as shear thinning in complex fluids, and current rectification in electrical transistors. Such behaviors become especially prevalent in nanoscale systems, where microscopic fluctuations play an important role. As a consequence, understanding the origin of nonlinear effects are key to the design and manipulation of nanoscale devices. One of the most efficient methods to compute nonlinear transport coefficient is direct nonequilibrium molecular simulation, where a current is driven through the system by application of specific boundary conditions \cite{tenenbaum1982stationary,baranyai1999steady}, or by altering the equations of motion \cite{hoover1980lennard,evans1982homogeneous,muller1997simple}. Such methods, however, are generally not transferable among different transport processes, and the result can be sensitive to how the current is generated \cite{tuckerman1997modified,tenney2010limitations}. Moreover, with direct simulation the computed response is not easily connected to specific molecular degrees of freedom.

Kubo in his original framework also derived a perturbation expansion for the nonequilibrium phase space distribution. However, such expression, as well as many that followed \cite{efremov1969fluctuation,stratonovich1970contribution}, have been difficult to translate into measurable forms with clear physical interpretations. Several non-perturbative treatments were proposed around the same time \cite{yamada1967nonlinear,bochkov1977general}, yet an explicit expression for computing nonequilibrium averages is lacking. The simplest generalization of the Green-Kubo formulas that fills in this gap is perhaps the transient time correlation function formalism \cite{visscher1974transport,dufty1979nonlinear,cohen1983kinetic}. It gives an exact relation between the nonequilibrium ensemble average of an observable and the transient time correlation function between the observable and the current. In practice, a nonequilibrium simulation has to be performed, yet the statistical uncertainty in the nonequilibrium average is often much larger than that of the direct nonequilibrium simulation \cite{evans1988transient}.

The last two decades have observed a growing interest in the field of stochastic thermodynamics, stimulated by new possibilities in experimental interrogation of nanoscale systems.
Nonequilibrium fluctuation theorems \cite{jarzynski1997nonequilibrium,crooks1999entropy,gallavotti1995dynamical} and thermodynamic uncertainty relations \cite{barato2015thermodynamic,gingrich2016dissipation} impose constraints on the distribution functions of fluctuating thermodynamic quantities like heat and work, which represent refinements of the second law to driven systems. These theoretical developments have enabled the establishment of a more formal theory for nonequilibrium steady states. In particular, a considerable amount of work has been devoted to predicting response coefficients for stochastic systems. Extended fluctuation-dissipation theorems have been derived for linear responses of nonequilibrium steady states 
\cite{harada2005equality,speck2006restoring,speck2009extended,baiesi2009fluctuations}, which can be translated into second order response around equilibrium \cite{basu2015nonequilibrium}. More generally, multivariate fluctuation relations imply connections between transport coefficients and cumulants of the current \cite{andrieux2004fluctuation,gaspard2013multivariate}.
Many of these advances have been enabled by large deviation theory \cite{ellis2007entropy,touchette2009large}, which provides a set of mathematical tools for characterizing and evaluating fluctuations in nonequilibrium systems. These tools have been employed widely to illustrative model systems, which not only serve as testing grounds for theories, but also reveal new insights into nonequilibrium phenomena \cite{hedges2009dynamic,hurtado2011spontaneous, limmer2014theory,pre2018current}.
 
In the present work, we aim to translate these earlier efforts to a larger class of nonequilibrium systems, and derive an expression for arbitrarily high order transport coefficients in terms of equilibrium correlation functions. To compute the correlation functions, we rely on the direct evaluation of large deviation functions (LDFs). We have shown in a previous work \cite{gao2017transport} that linear transport coefficients can be extracted from the second derivative of LDFs in a statistically efficient way. Here, we extend the methodology to nonlinear transport regimes. In what follows, we first formulate our theory
and then demonstrate our method in a collection of examples, including a Brownian ratchet, a model thermal rectifier, and a tracer particle in a convective flow.

\section*{Theory and Methodology}
We are interested in stochastic systems maintained in nonequilibrium steady states by an affinity $X$, which can either be an external force, or contacts with different thermodynamic reservoirs. We assume that the macroscopic current $\mathcal{J}$ generated in the system as a response to $X$ can be written as a polynomial expansion,
\begin{equation}
\label{Eq:Li}
\mathcal{J} = L_0+L_1 X+L_2 X^2+L_3 X^3+\cdots,
\end{equation}
where $L_0$ is the current in the absence of the affinity, which should vanish for equilibrium systems. $L_1$ is known as the linear transport coefficient, and all the higher order coefficients $L_2,L_3,\cdots$, are nonlinear transport coefficients. The generality of this polynomial expansion will be discussed towards the end of this paper.
Given this assumption, in this section we derive an explicit expression for these transport coefficients. Although we will restrict ourselves to a single type of affinity, the generalization to multiple ones is straightforward.

\subsection*{Derivation of nonlinear coefficients from LDFs} 
We consider a continuous stochastic trajectory of length $t_N$ denoted by $\tilde{x}$, where $\tilde{x}_t=\{\bm{r}(t),\bm{v}(t) \}$ is the specific configuration of the system at time $t$, with coordinates $\bm{r}(t)$ and  velocities $\bm{v}(t)$. We define a relative stochastic action $\beta U[\tilde{x}]$ by a ratio of the probability of observing a specific path with and without the affinity, 
\begin{equation}
\frac{P_X[\tilde{x}]}{P_0[\tilde{x}]}=e^{\beta U[\tilde{x}]},
\end{equation}
where the subscripts denote the value of the affinity; $\beta=1/k_B T$ where $T$ is the temperature, $k_B$ is Boltzmann constant that will be set to 1 in the following calculations.

We next decompose the total relative path action into two parts according to their time reversal symmetry,
\begin{equation}
U[\tilde{x}]=A[\tilde{x}]+S[\tilde{x}],
\end{equation}
with $A[\tilde{x}]$ denoting the asymmetric part, and $S[\tilde{x}]$ the symmetric part. Specifically, if we define an operator $\mathbb{T}$ which returns the time-reversed counterpart of a path $\tilde{x}$ so that $\mathbb{T}\tilde{x}_t=\{\bm{r}(t_N-t),-\bm{v}(t_N-t) \}$, it follows that $A[\mathbb{T}\tilde{x}]=-A[\tilde{x}]$, $S[\mathbb{T}\tilde{x}]=S[\tilde{x}]$.
The time asymmetric part is the entropy production of the irreversible process, which can be written as the product of the affinity and the conjugated time extensive current \cite{onsager1931reciprocal},
\begin{equation}
A[\tilde{x}]=t_NJX,\hspace{2mm}J[\tilde{x}]=\frac{1}{t_N}\int_{0}^{t_N}dt\hspace{2mm}j(\tilde{x}_t)
\end{equation}
The symmetric part is often referred to as the activity \cite{maes2008steady}, 
\begin{equation}
S[\tilde{x}]=\sum_{i=1}^p t_NQ_iX^i,\hspace{2mm}Q_i[\tilde{x}]=\frac{1}{t_N}\int_{0}^{t_N}dt\hspace{2mm}q_i(\tilde{x}_t),
\end{equation}
The number of terms $p$ depends on system details, though for Gaussian processes it is often only 1. For details, we show the derivation of $J$ and $Q_i$'s for a general underdamped Langevin system in the Appendix.

Next, we construct a LDF of the form 
\begin{equation}
\label{Eq:LDF}
\psi_X(\boldsymbol{\lambda})
=\lim_{t_N\rightarrow\infty}\frac{1}{t_N}\ln\left \langle e^{-(\lambda_J t_N J+\sum_i\lambda_{Q_i} t_N Q_i)} \right \rangle_X,
\end{equation}
where $\boldsymbol{\lambda}=(\lambda_J,\lambda_{Q_1},\lambda_{Q_2},\cdots)$, i.e. each dynamical variable is exponentially biased by a conjugated $\lambda$. The bracket denotes path ensemble average with respect to the path probability density $P_X[\tilde{x}]$. It follows by definition that
\begin{equation}
\begin{aligned}
\label{Eq:link}
\psi_X(\lambda_J,&\lambda_{Q_1},\lambda_{Q_2},\cdots)\\
&=\psi_0(\lambda_J-\beta X, \lambda_{Q_1}-\beta X,\lambda_{Q_2}-\beta X^2,\cdots),
\end{aligned}
\end{equation}
which provides a symmetry that links the statistical bias ${\boldsymbol{\lambda}}$ to the physical driving $X$. This relation is distinct from a fluctuation theorem and encodes the fact that $X$ is not a conjugate variable to $J$, but rather a linear combination of $J$ and the $Q_i$'s. The derivatives of the cumulant generating function provide information about the self and cross correlation functions of $J$ and the $Q_i$'s. For example, the nonequilibrium average current is given by the first derivative
\begin{equation}
\label{Eq:derivative}
\left \langle J \right \rangle_X=-\frac{\partial \psi_X({\boldsymbol{\lambda}} )}{\partial \lambda_J}\Bigr|_{\substack{\boldsymbol{\lambda}=\mathbf{0}}}=-\frac{\partial \psi_0({\boldsymbol{\lambda}} )}{\partial \lambda_J}\Bigr|_{\substack{\lambda_J=-\beta X,\lambda_{Q_i}=-\beta X^i}}.
\end{equation}
Expanding $\psi_0({\boldsymbol{\lambda}} )$ assuming $\beta X$ is small,
\begin{equation}
\begin{aligned}
\label{Eq:JofX}
\left \langle J \right \rangle_X&=\left \langle J \right \rangle_0+\beta t_N \left [\left \langle (\delta J)^2 \right \rangle_0+\left \langle \delta J\delta Q_1 \right \rangle_0 \right ]X\\
&+\frac{\beta^2t_N^2}{2}\left[\left \langle (\delta J)^3 \right \rangle_0 +\left \langle \delta J(\delta Q_1)^2 \right \rangle_0+2\left \langle (\delta J)^2\delta Q_1 \right \rangle_0 \right.\\
&\left.+\frac{2}{\beta t_N} \left \langle \delta J\delta Q_2 \right \rangle_0 \right ]X^2+\cdots,
\end{aligned}
\end{equation}
we find a microscopic relation between the average current and affinity, where for brevity, we have only written down explicitly terms up to $O(X^2)$. The notation $\left \langle A_1A_2\cdots A_n \right \rangle=t_N^{-n}\int_0^{t_N}d{\bf t}^n \left \langle a_1(\tilde{x}_{t_1})a_2(\tilde{x}_{t_2})\cdots a_n(\tilde{x}_{t_n} )\right \rangle$ and $\delta A=A-\left\langle A\right\rangle_0$ are adopted throughout. By comparing with Eq.~\ref{Eq:Li}, and assuming the time averaged current is equal to the macroscopic current, the expansion in Eq.~\ref{Eq:JofX} provides explicit expressions for arbitrarily high order transport coefficients in terms of multi-time correlation functions in the absence of the affinity, which can be computed from $\psi_0({\boldsymbol{\lambda}})$. 
While multi-time correlation functions are in general difficult to converge by direct evaluation, $\psi_0({\boldsymbol{\lambda}})$ can be computed efficiently using importance sampling methods \cite{nemoto2014computation,nemoto2017finite,klymko2018rare,ray2018exact}. 

To highlight the novelty of our method, we discuss how our expression is different from a few previous results in the literature. Firstly, our result is intimately related to work by Maes et al \cite{colangeli2011meaningful, basu2015frenetic}, though our current expression Eq.~\ref{Eq:JofX} is different beyond the third order. This stems from the fact that the current is derived in terms of a cumulant expansion instead of a moment expansion, as seen clearly in Eq.~\ref{Eq:JofX}, and from rewriting Eq.~\ref{Eq:derivative},
\begin{equation}
\left \langle J \right \rangle_X=\frac{\left\langle J e^{\beta U}\right\rangle_0}{\left\langle e^{\beta U}\right\rangle_0}.
\end{equation}
The cumulant expansion ensures better convergence for the transport coefficients, especially for dynamical variables whose distributions are close to Gaussian \cite{freed1968generalized,wang1998application}.
Secondly, our method is distinct from the multivariate fluctuation relations that have been derived before by constructing the LDF of only current observables \cite{gaspard2013multivariate}, where transport coefficients are expressed as mixed derivatives of both $\lambda$'s and the affinities. Here, by introducing the time symmetric observables $Q_i$'s, the knowledge of the equilibrium function $\psi_0({\boldsymbol{\lambda}} )$ completely determines the current-affinity relationship. 

\subsection*{Simplification by symmetries} 
Decomposing the total action by time reversal symmetry enables us to greatly simplify Eq.~\ref{Eq:JofX}. If the reference system without the affinity is in equilibrium, then it obeys microscopic time reversibility $P_0[\mathbb{T}\tilde{x}]=P_0[\tilde{x}]$. It follows that the time reversal odd terms, such as the average current $\left \langle J \right \rangle_0$, and terms like $\left \langle JQ_i \right \rangle_0$, $\left \langle JQ_i^2 \right \rangle_0$ and $\left \langle J^3 \right \rangle_0$, will vanish. This reduces Eq.~\ref{Eq:JofX} to
\begin{equation}
\label{Eq:JofXS}
\left \langle J \right \rangle_X=\beta t_N\left \langle J^2 \right \rangle_0 X+\beta^2t_N^2\left \langle J^2Q_1 \right \rangle_0X^2+O(X^3).
\end{equation}
Note that the linear response reduces to the normal fluctuation-dissipation theorem, while higher order responses are described by higher order correlations between $J$ and $Q_i$'s. These correlation functions are given by derivatives of $\psi_0(\boldsymbol{\lambda})$. Specifically, the linear and first nonlinear transport coefficients are
\begin{equation}
\label{Eq:Ls}
L_1 =  \beta  \frac{\partial^2 \psi_0}{\partial \lambda^2_J}\Bigr|_{\substack{\boldsymbol{\lambda}=\mathbf{0}}} \, ,\quad  L_2 =  -\beta^2 \frac{\partial^3 \psi_0}{\partial \lambda_J^2 \partial \lambda_{Q_1}}\Bigr|_{\substack{\boldsymbol{\lambda}=\mathbf{0}}}.
\end{equation}
Further simplifications can be made if the system exhibits certain spatial symmetries so that the dynamics is unchanged upon inverting the coordinates along a specific axis. 

\begin{figure}[t]
\centering
\includegraphics[width=.95\linewidth]{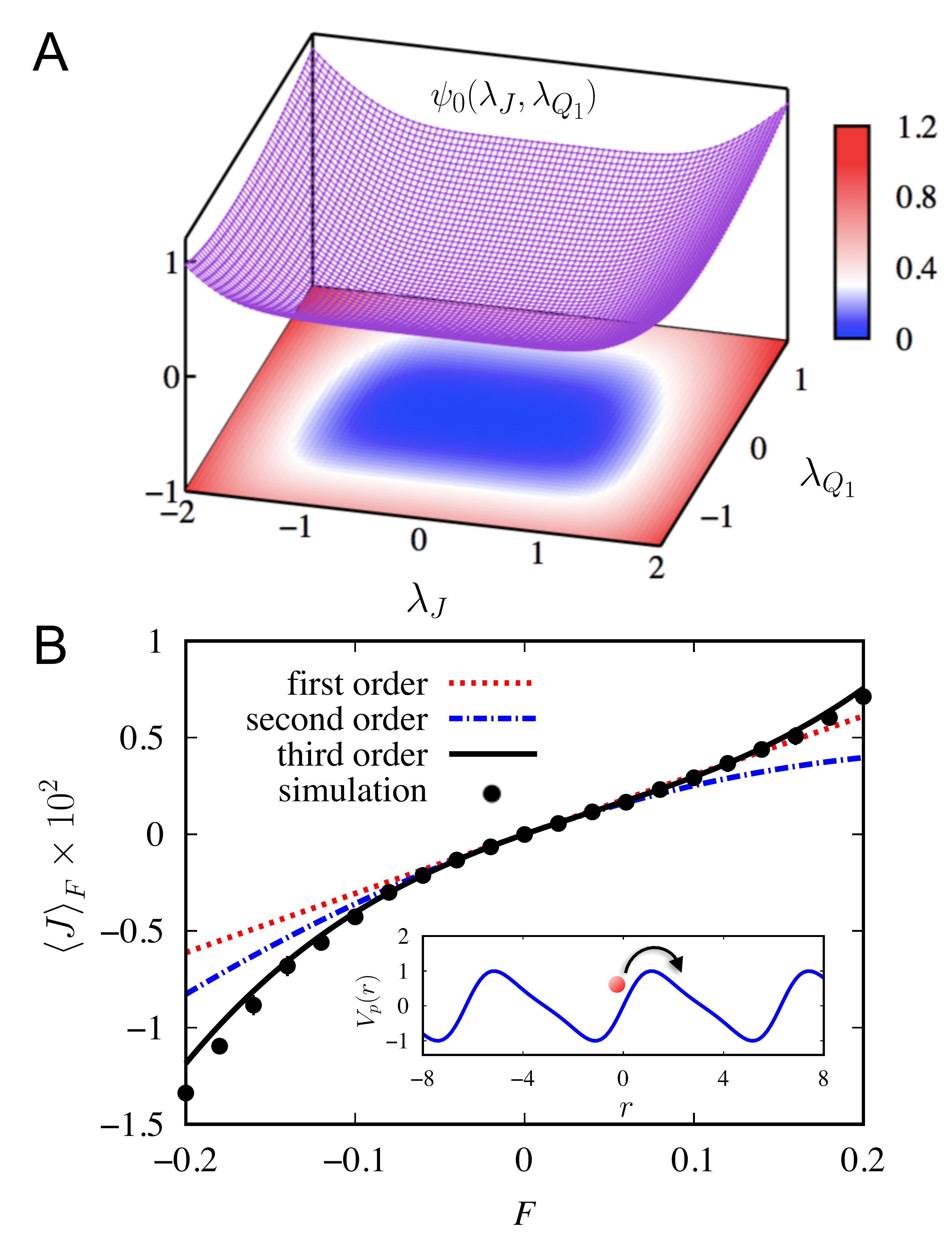}
\caption{\label{fig:ratchet} Computation of nonlinear transport coefficients of a Brownian ratchet. The predicted coefficients are derived from the 2D surface of the LDF in $A$, where a contour plot is projected onto the $\lambda_J-\lambda_{Q_1}$ plane. $B$ shows the comparison between nonequilibrium simulation results (black dots) and theoretical approximations for the current up to first (dotted red), second (dotted dashed blue), and third (solid black) order. (\textit{Inset}) The potential landscape of the Brownian ratchet. }
\end{figure}

\subsection*{Numerical evaluation of large deviation functions}
To evaluate the large deviation function defined in Eq.~\ref{Eq:LDF}, we use a diffusion Monte Carlo method known as the cloning algorithm \cite{giardina2006direct}, where an ensemble of trajectories is propagated in parallel. Each individual trajectory is known as a walker, and collectively they undergo a population dynamics that samples a biased trajectory ensemble.
A more detailed description of the algorithm can be found in our earlier work \cite{gao2017transport}.

One of the key factors in determining the statistical efficiency of the cloning algorithm is the number of independent trajectories sampled over $t_N$. In all the calculations shown in this paper, walker numbers are chosen individually for each parameter so that at the end of the simulation, the number of independent walkers which have not been replaced is at least on the order of $10^2$. 

While transport coefficients can also be computed directly through nonequilibium simulation methods, we note that by evaluating them from equilibrium fluctuations, our method is less sensitive to finite size effects arising from boundary conditions and altered equations of motion required to simulate a driven system.
In addition, while direct simulation measures the nonequilibrium response at a finite value of the affinity, our method generically generates the response for a continuum of affinities. Indeed, if one is specifically interested in the current response at a finite field $X$, it suffices to measure the local curvature of $\psi_X$ at $\boldsymbol{\lambda}=\boldsymbol{0}$ by computing the equilibrium large deviation function $\psi_0$ around $\lambda_J=\beta X, \lambda_{Q_i}=\beta X^i$ according to Eq.~\ref{Eq:link}, where we have established a direct connection between the statistical biased and nonequilibrium ensembles.

\section*{Models}
\subsection*{Brownian Ratchet} 
To validate our method, we first consider a single Brownian particle moving on a one-dimensional asymmetric potential landscape $V_p(r)=\sin(r/l+\sin(r/l)/2)$.  This is a simplest continuous model that exhibits an asymmetric response of the particle's displacement to an additional constant force. The dynamics of the particle with unit mass are described by the overdamped Langevin equation
\begin{equation}
\gamma \dot{r}=-\frac{d V_p(r)}{d r}+F+\eta.
\end{equation}
where $F$ is a constant force, $\gamma$ is the friction and $\eta$ is the white noise satisfying 
$\left \langle \eta(t) \right \rangle=0,
\left \langle \eta(0)\eta(t) \right \rangle=2\gamma \delta(t)/\beta$. We set $\gamma$ and $l$ to 1, to define a reduced unit system and take $\beta=1/0.3$. Considering the mobility of the particle in response to the affinity $X=F/2$, the dynamical variables take the form
\begin{equation}
j(\tilde{x}_t)=\dot{r}(t),\hspace{0.2cm} q_1(\tilde{x}_t)=\gamma^{-1} \frac{d V_p}{d r}(t).
\end{equation}
Here, the time symmetric part is characterized by the force exerted onto the particle by the ratchet potential. There is an additional second order term $q_2(\tilde{x}_t)$, which is a constant and does not enter into the expression for the current.
Expanding Eq.~\ref{Eq:JofX} up to the third order, we arrive at an expression for the integrated nonequilibrium current, or displacement,
\begin{equation}
\begin{aligned}
\label{Eq:JFull}
\left \langle J \right \rangle_F&=\beta t_N\left \langle J^2 \right \rangle_0 \frac{F}{2} +\beta^2 t_N^2\left \langle J^2Q_1 \right \rangle_0 \frac{F^2}{4} \\ 
&+\beta^3 t_N^3  \left [\frac{1}{6}(\left \langle J^4 \right \rangle_0 - 3 \left \langle J^2 \right \rangle_0^2)+ \right .\\
&\left .\frac{1}{2}(\left \langle J^2Q_1^2 \right \rangle_0-\left \langle J^2 \right \rangle_0\left \langle Q_1^2 \right \rangle_0)  \right ] \frac{F^3}{8}+O(F^4).
\end{aligned}
\end{equation}
It is worth noting that the higher order responses are described not only by the current fluctuations, but also the correlations between current and force fluctuations. If the potential is symmetric, the second order term will vanish due to inversion symmetry. It is the asymmetry in the ratchet potential, and as a consequence, the asymmetry in the forces, that gives rise to the even order terms. Specifically, the correlation between the squared current and the force dictates the size of the rectification of the ratchet. 

In this simple case, the LDF $\psi_0({\boldsymbol{\lambda}})$ can be computed numerically exactly by diagonalizing the tilted generator \cite{touchette2009large},
\begin{equation}
L_{\boldsymbol{\lambda}}=-\frac{\partial V_p}{\partial x} \left (\frac{\partial}{\partial x}-\lambda_J \right )+T \left (\frac{\partial}{\partial x}-\lambda_J \right )^2-\lambda_{Q_1}\frac{\partial V_p}{\partial x},
\end{equation}
the largest eigenvalue of which is the LDF. To solve for the eigenvalues, we construct the tilted operator with a normalized Fourier basis set $\exp({ikx})$ where $k\in [-15,15]$ and is an integer. The 2D surface of $\psi_0({\boldsymbol{\lambda}})$ is shown in Fig. \ref{fig:ratchet}$A$. The obvious deviation from the normal distribution, especially the asymmetry in the $\lambda_{Q_1}$ direction, leads to the observed nonlinear behavior of $\langle J \rangle_F$. The correlation functions appearing in the average current are computed from numerical derivatives of $\psi_0$ as in Eq.~\ref{Eq:Ls}, yielding $t_N\left \langle J^2 \right \rangle_0=0.0184$, $t_N^2\left \langle J^2Q_1 \right \rangle_0=-0.0194$, $t_N^3(\left \langle J^4 \right \rangle_0-3\left \langle J^2 \right \rangle_0^2)=0.712$, and $t_N^3(\left \langle J^2Q_1^2 \right \rangle_0-\left \langle J^2 \right \rangle_0\left \langle Q_1^2 \right \rangle_0)=-0.033$. Fitting errors in the coefficients are negligible. The nonequilibrium simulation results are obtained by integrating the equation of motion using a second-order Runge-Kutta algorithm \cite{honeycutt1992stochastic} with a time step $h=10^{-3}$. Numerical results shown are averaged over $10^3$ realizations with a total observation time $t_N=10^3$. In Fig. \ref{fig:ratchet}$B$, we compare direct nonequilibrium simulation results with our theoretical predictions, which improve as we include higher order terms. It is worth noting that in this case, as the large deviation function can be solved conveniently by matrix diagonalization, the computational efficiency of our method is far superior compared to direct simulation, which requires sampling of the nonequilibrium dynamics.

\subsection*{Model Thermal Rectifier}

We next turn to an interacting system, a 1D thermal rectifier.
A thermal rectifier is a type of material with intrinsic structural asymmetry so that it exhibits 
an asymmetric heat transport response when a temperature gradient is applied.
Here we model a 1D thermal rectifier using the linearly mass-graded Fermi-Pasta-Ulam-Tsingou (FPUT) chain, which has been shown to capture thermal rectification behavior \cite{yang2007thermal}. The mass of the $i$th particle is $m_i=m_\mathrm{min}+(i-1)(m_\mathrm{max}-m_\mathrm{min})/(N-1),\hspace{2mm} i=1,\cdots,N$,   
with $m_\mathrm{max}=20m_\mathrm{min}$, and $N=20$ is the total number of particles. The particles oscillate around their average position $b_i=ia$, while interacting with neighboring particles through the quartic FPUT potential, $V_{\mathrm{FPUT}}(\bm{r})=\sum_{i=1}^{N+1}\kappa (r_i-r_{i-1}-a)^2/2+c(r_i-r_{i-1}-a)^4/4,$
where $\kappa$ and $c$ are the harmonic and anharmonic coupling, and $a$ is the lattice constant. Fixed boundary conditions are applied by fixing two fictitious particles at $b_0$ and $b_{N+1}$. We set $m_\mathrm{min}=\kappa=a/2=1$, to define a dimensionless unit system, in which we also let $c=1$. The equation of motion is integrated by the velocity Verlet algorithm with a timestep of $h=5\times 10^{-3}$. 

\begin{figure}[t]
\centering
\includegraphics[width=.9\linewidth]{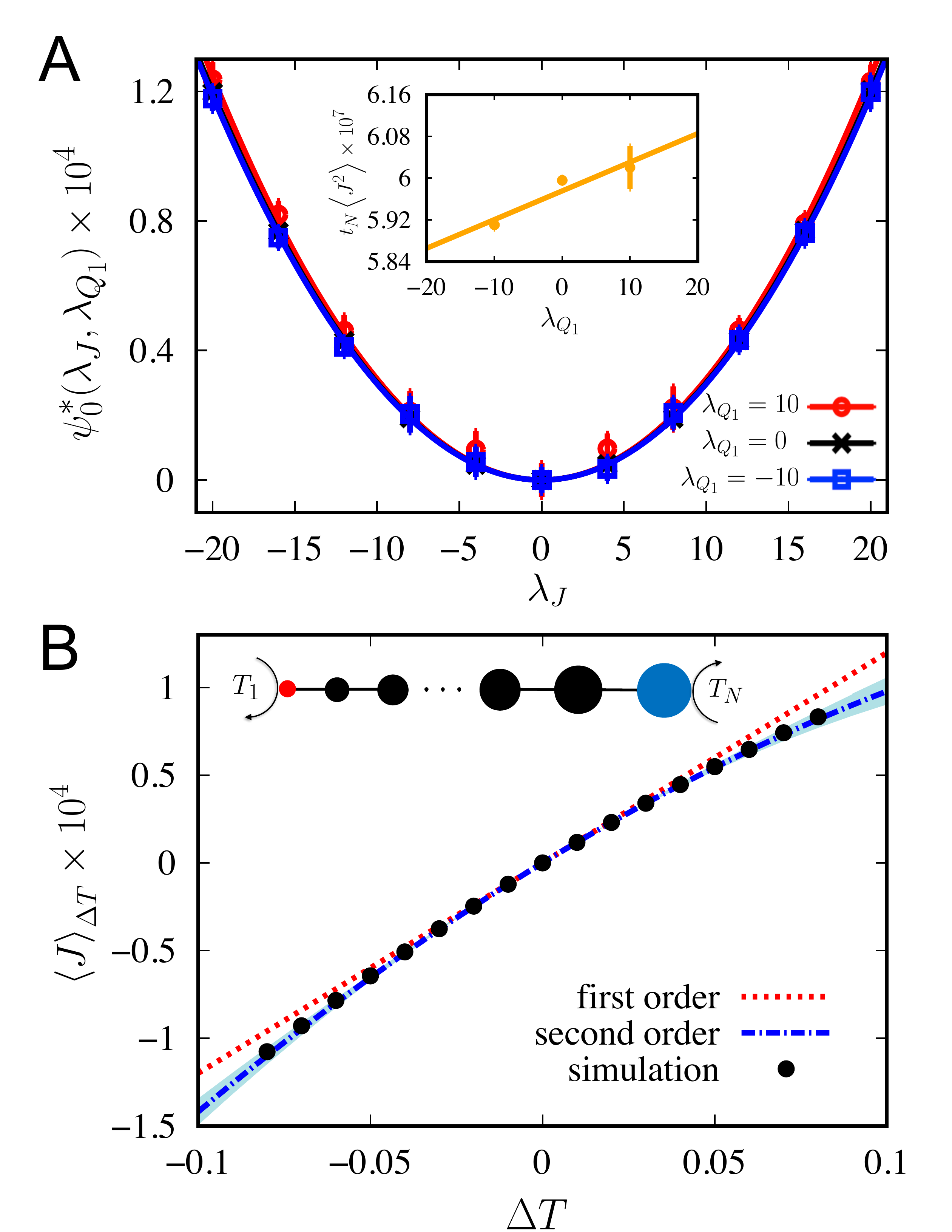}
\caption{\label{fig:FPU} Current rectification in 1D mass-graded FPU chain. $A$ shows the shifted LDF $\psi^*_0(\lambda_J,\lambda_{Q_1})$ at $\lambda_{Q_1}=10$ (red), $0$ (black) and $-10$ (blue). Solid lines are quadratic fits. (\textit{Inset}) Variance of the current as a function of $\lambda_{Q_1}$, measured from the curvature of the LDFs. $B$ shows the nonequilibrium current (black dots) and the prediction from Eq. 17 approximated to the first (dotted red) and second order (dotted dashed blue), where the shaded area represents the statistical error in second order coefficient. (\textit{Inset}) Schematic of the simulated system.}
\end{figure}

To apply an external temperature gradient, we place the particles on the two ends in contact with infinitely large thermal reservoirs kept at temperatures $T_1=T_0+\Delta T$ and $T_N=T_0-\Delta T$, respectively, where the average temperature $T_0=0.1$. This is implemented by an Andersen thermostat \cite{frenkel2001understanding} on each of the two end particles, where the time interval $\Delta t$ between successive collisions are distributed as $P(\Delta t)=\Gamma e^{-\Gamma \Delta t}$
and $\Gamma = 0.8$ is the coupling strength to the bath. The affinity for the heat transfer process is $
X=1/(T_0-\Delta T)-1/(T_0+\Delta T)\approx 2\Delta T/T_0^2$. In the steady state limit, the heat current from the two ends must be same in magnitude but opposite in direction, so we define the heat current as 
\begin{equation}
j(\tilde{x}_t) = \frac{T_0}{4 }\left (m_1\dot{v}_1v_1-v_1 F_1^{\mathrm{FPUT}}-m_N\dot{v}_Nv_N+v_NF_N^{\mathrm{FPUT}} \right ),
\end{equation}
where $F_i^{\mathrm{FPUT}}=-\partial V_{\mathrm{FPUT}}(\bm{r})/\partial r_i$. This can be interpreted as the rate of change in the kinetic and potential energy, averaged over contributions from particle $1$ and $N$, scaled by a factor of $T_0/2$. This constant factor is essential in matching our definition of the current. 

For the Andersen thermostat, the path probability can be written as
\begin{equation}
\begin{aligned}
P_{\Delta T}[\tilde{x}]&\propto\prod_{k}\sqrt{\frac{m_1}{2\pi T_1}}\exp[-\frac{m_1v_1(t_k)^2}{2T_1}]\\
&\times\prod_{l}\sqrt{\frac{m_N}{2\pi T_N}}\exp[-\frac{m_Nv_N(t_l)^2}{2T_N}]
\end{aligned}
\end{equation}
where $t_k$($t_l$) are times at which the first(last) particle collides with the bath.
From this we can derive the relative path action, and the time-symmetric part follows as
\begin{equation}
Q_1[\tilde{x}]=\frac{T_0}{4t_N} \left (\sum_{k}m_1v_1^2(t_k)-\sum_{l}m_Nv_N^2(t_l) \right )-J[\tilde{x}] \, .
\end{equation}
Note that since we have chosen to express the action in terms a Taylor expansion of $\Delta T$, we have an infinite series of time-symmetric parts; however, we could have formulated in terms of $\Delta(1/T)$, in which case there will only be one single term $Q_1$.
The average current as a function of $\Delta T$ becomes
\begin{equation}
\begin{aligned}
\left \langle J \right \rangle_{\Delta T}&=\beta t_N\left \langle J^2 \right \rangle_0\frac{2\Delta T}{T_0^2}\\
&+ \beta^2  t_N^2\left \langle J^2Q_1 \right \rangle_0 \left (\frac{2 \Delta T}{T_0^2} \right)^2+O(\Delta T^3)
\end{aligned}
\end{equation}
The first term is the standard Green-Kubo result for the thermal conductivity in terms of an integrated heat flux autocorrelation function. The second term correlates the squared heat flux with the instantaneous temperature difference, and results in an asymmetric response of the current. Since the masses of the two ends are different, the time dependent temperature fluctuations on either side of the chain need not be the same. This expression illustrates that thermal current rectification is a product of microscopic correlations between instantaneous temperature gradients and heat fluxes.

The evaluation of $\psi_0({\boldsymbol{\lambda}})$ from diffusion Monte Carlo is shown in Fig. \ref{fig:FPU}$A$, while the nonequilibrium simulation results of the current is shown in Fig. \ref{fig:FPU}$B$. Both the LDFs and the nonequilibrium simulation results averaged over 9600 realizations are calculated for trajectories with $t_N=2\times 10^5$. The LDFs are evaluated at $\lambda_J\in[-32,32]$ for $\lambda_{Q_1}=10$, and $\lambda_J\in[-20,20]$ for $\lambda_{Q_1}=0,-10$. Eight independent samples are calculated at each combination of $\boldsymbol{\lambda}$, and standard deviations are plotted as error bars in Fig. \ref{fig:FPU}$A$. For each set of samples at a specific $\lambda_{Q_1}$, a parabola is fit, and the error bars in inset of Fig. \ref{fig:FPU}$A$ are the standard deviation among the 8 curvatures. To estimate the statistical error in the second order transport coefficient, the fitted slope in the inset of Fig. \ref{fig:FPU}$A$ is evaluated individually for each of the 8 sample sets, and standard error of the mean is reported in Fig. \ref{fig:FPU}$B$. While $t_N\left \langle J^2 \right \rangle_0$ is measured by the curvature of $\psi_0$ with $\lambda_{Q_1}=0$, $t_N^2\left \langle J^2Q_1 \right \rangle_0$ is measured by how the curvature changes as we change $\lambda_{Q_1}$. Given the definition of $Q_1$, the change of the curvature with $\lambda_{Q_1}$ directly reports the change in the thermal conductivity with a temperature gradient.  To make a better comparison, we plot the shifted LDFs $\psi_0^*(\lambda_J,\lambda_{Q_1})=\psi_0(\lambda_J,\lambda_{Q_1})-\psi_0(\lambda_J,\lambda_{Q_1}=0)$ so that all the curves have the same minimum value at $\lambda_J=0$. Our method correctly predicts the rectification behavior with high accuracy.

\begin{figure}[t]
\centering
\includegraphics[width=.9\linewidth]{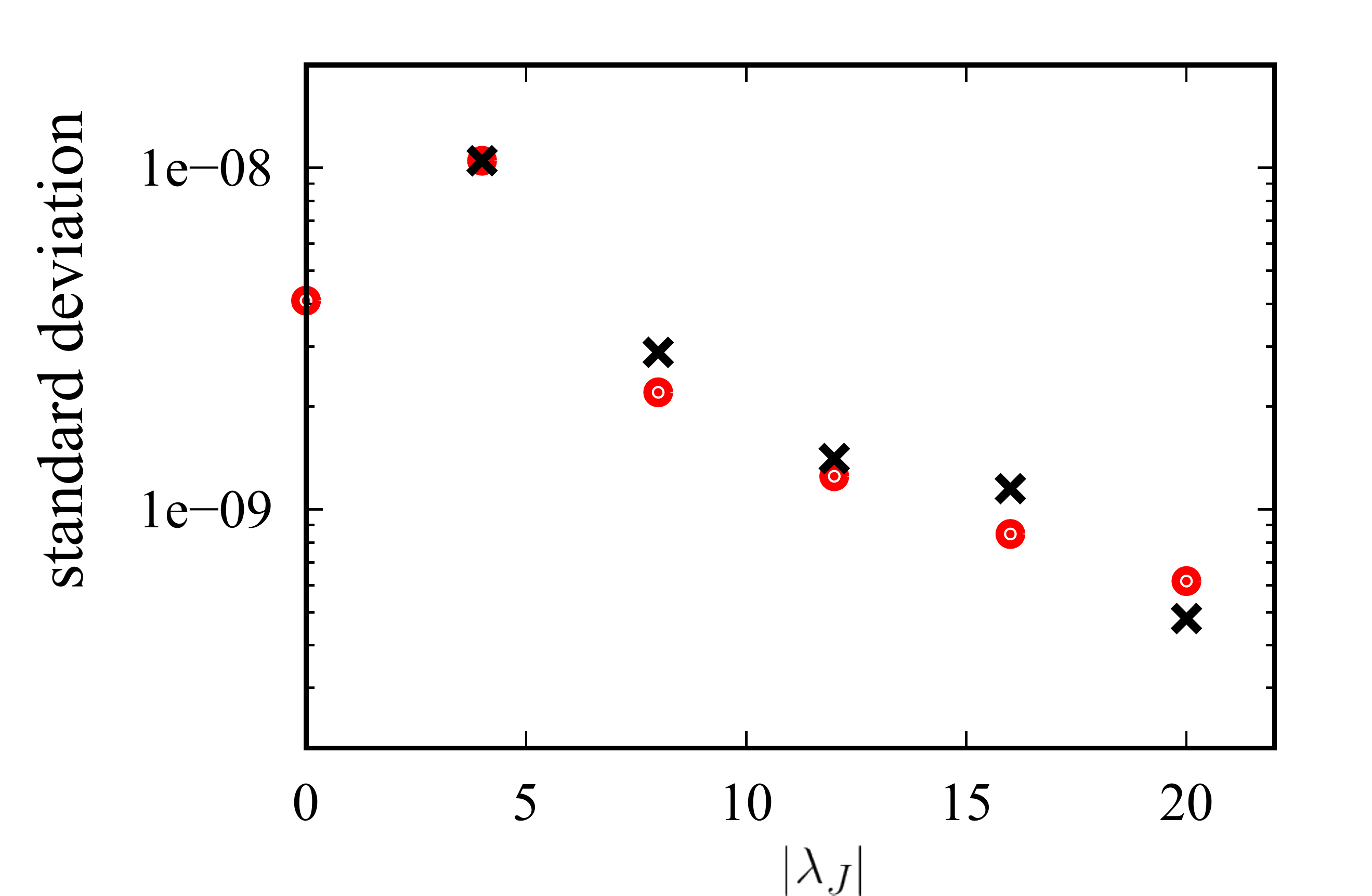}
\caption{\label{fig:staterror} Comparison of statistical error in $t_N^2\left \langle J^2Q_1 \right \rangle_0$ from direct evaluation (plotted as $\lambda_J=0$) and the cloning algorithm for different $\lambda_J$'s while fixing $\lambda_{Q_1}=10$. Red dots are for positive $\lambda_J$'s while black crosses are for negative ones. }
\end{figure}

To demonstrate the statistical efficiency of our approach, in Fig. \ref{fig:staterror} we compare the statistical error in the evaluation of $t_N^2\left \langle J^2Q_1 \right \rangle_0$ by direct evaluation and using the cloning algorithm. The direct evaluation, plotted as $\lambda_J=0$ in Fig. \ref{fig:staterror}, is evaluated by computing the triple correlation function by brute force averaged among $1.2\times 10^5$ independent trajectories. In the cloning algorithm, we evaluate $\psi(\lambda_J,\lambda_{Q_1}=10)$ for various $\lambda_J$'s with the same number of walkers $N_w=1.2\times 10^5$, and compute the correlation function by $t_N^2\left \langle J^2Q_1 \right \rangle_0=2[\psi_0(\lambda_J,\lambda_{Q_1})-\psi_0(\lambda_J,0)-\psi_0(0,\lambda_{Q_1})]/(\lambda_J^2\lambda_{Q_1})$. Statistical errors are estimated by the standard deviation from 9 independent simulations. 
We have chosen the same number of walkers and independent trajectories to ensure that the computational effort in terms of number of integration steps is the same. For equal computational effort evaluating $\left \langle J^2Q_1 \right \rangle_0$ from the large deviation function exhibits smaller statistical error for sufficiently large $\lambda_J$'s, as shown in Fig. \ref{fig:staterror}. For a fixed $\lambda_{Q_1}$, the error in the curvature of $\psi_0$ with respect to $\lambda_J$ should scale as $\propto\sqrt{\psi_0''(\lambda_J)/\tilde{N}_w}/\psi_0(\lambda_J)=1/(\lambda_J^2\sqrt{\tilde{N}_w})$, where $\tilde{N}_w$ is the number of uncorrelated walkers. Even though the correlation between the walkers as $\lambda_J$ increases brings in a non-trivial dependence of $\tilde{N}_w$ on $\lambda_J$, overall the cloning algorithm still out-performs the direct evaluation for $|\lambda_J|>8$ by about an order of magnitude.

\subsection*{Tracer particle in a convective flow} 

In the last section, we discuss a system out of equilibrium even at $X=0$. This is fundamentally different from the two cases above, as even at linear response traditional Green-Kubo formulas do not apply. Nevertheless the response of the system is still encoded in a LDF computed at $X=0$, though one evaluated in the nonequilibrium steady state. The previously derived extended fluctuation-dissipation theorems \cite{harada2005equality,speck2006restoring,speck2009extended} follow naturally from our general expression for the nonequilibrium current in Eq.~\ref{Eq:JofX}.

We consider an underdamped particle with unit mass moving in a 2D hydrodynamic flow 
\begin{equation}
\dot{v}_x=-\gamma(v_x-U_x)+F+\eta_x,\hspace{2mm}\dot{v}_y=-\gamma(v_y-U_y)+\eta_y,
\end{equation}
where $U_x=\partial\phi(\bm{r})/\partial r_y$ and $U_y=-\partial\phi(\bm{r})/\partial r_x$ describe the divergenceless flow of the stream function $\phi(\bm{r})=LU_0\sin(2\pi r_x/L)\sin  (2\pi r_y/L)/(2\pi)$. We set $U_0=L=1$, which sets a scale for length and time. The noise $\eta_i$'s satisfy $\left \langle \eta_i(t) \right \rangle=0$, $\left \langle \eta_i(0)\eta_j(t) \right \rangle=2\gamma \delta_{ij}\delta(t)/\beta,i=x,y$. When the constant force $F=0$, it is the non-gradient form of the velocity field that forces the system out of equilibrium, though the average particle current still vanishes due to spatial symmetry.  As a consequence, the linear response of the particle current includes two terms proportional to $X$, as shown in Eq.~ \ref{Eq:JofX}. The extended fluctuation-dissipation theorem includes a term that is the correlation between the current and a time symmetric variable derived from the first order part of the relative action as has been described previously \cite{baiesi2009fluctuations}. Defining the affinity $X=F/2$, the dynamical variables are
\begin{equation}
j(\tilde{x}_t)=v_x(t),\hspace{0.5cm}q_1(\tilde{x}_t)=\dot{v}_x(t)/\gamma- U_x(t).
\end{equation}
The time symmetric term includes the particle's inertia, relative to its local flow velocity, which is only a function of the particle's position. The linear response of the current can be viewed as a Green-Kubo relation, but correlating particle's velocity relative to the velocity field $U_x$ \cite{harada2005equality,speck2006restoring}, up to a boundary term from the integral of $\dot{v}_x$. 

We note that the dynamical variables satisfy the relation $j(\tilde{x}_t)+q_1(\tilde{x}_t)=\eta_x/\gamma$ when $F=0$, which implies that $t_N^{n-1}\left \langle (J+Q_1)^n \right \rangle_0=(2/(\gamma\beta))^{n/2}(n-1)!!$ if $n$ is even, and vanishes if $n$ is odd. Note that this type of relationship is not restricted to this specific model - it is quite general for stochastic systems with quadratic path actions. These constraints on the moments reduce the number of unknown moments we have to compute. Here, we use the fact that the second order moments satisfy
$t_N\left \langle (J+Q_1)^2 \right \rangle_0=2/(\gamma\beta)$,
which allows us to rewrite the linear response as
\begin{equation}
\label{Eq:LRSS2}
\left \langle J \right \rangle_F=\beta  \left  (\frac{1}{\gamma\beta}+\frac{t_N}{2}\left \langle J^2 \right \rangle_0-\frac{t_N}{2}\left \langle Q_1^2 \right \rangle_0 \right )\frac{F}{2}+O(F^2).
\end{equation}
To compute the linear response coefficient, all we need are the curvatures of $\psi_0({\boldsymbol{\lambda}})$ along $\lambda_J=0$ and $\lambda_{Q_1}=0$, which are easily computable from diffusion Monte Carlo~\cite{giardina2006direct,gao2017transport}. 

\begin{figure}[t]
\centering
\includegraphics[width=.95\linewidth]{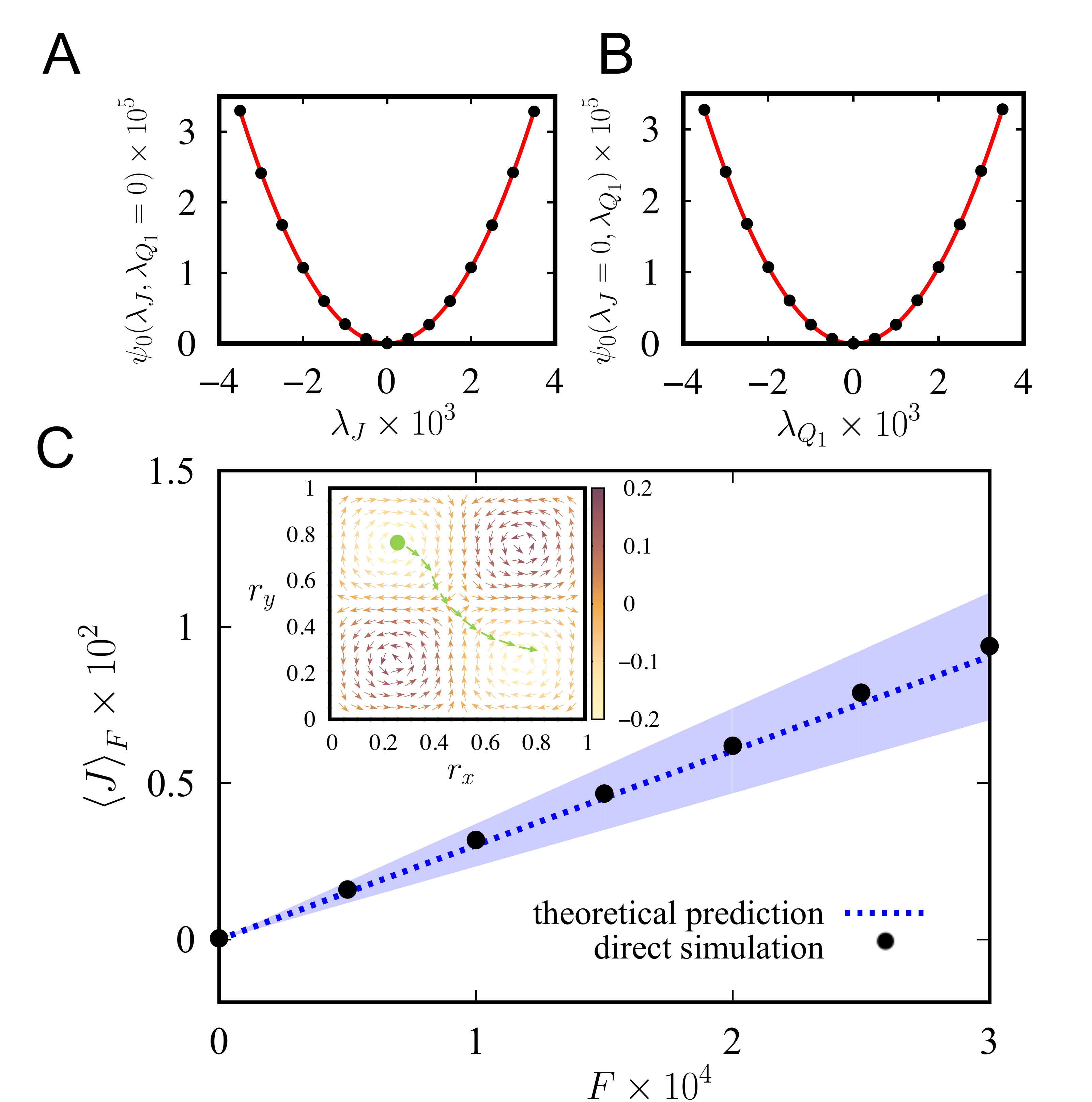}
\caption{\label{fig:tracer} Linear response of a tracer particle in a hydrodynamic flow.  LDFs are plotted as a function of $\lambda_J$ while fixing $\lambda_{Q_1}=0$ in $A$, and as a function of $\lambda_{Q_1}$ while fixing $\lambda_J=0$ in $B$. Red curves are quadratic fits. A comparison between direct nonequilibrium simulation results and the predicted response is shown in $C$. Shaded area is the statistical error of the theoretically predicted response. (\textit{Inset}) The flow field and a schematic trajectory of the tracer. Orange arrows illustrate the underlying velocity field and colors map the value of the stream function.}
\end{figure}

In the following calculations, we set $\gamma=0.1$, $\beta=0.5\times 10^{4}$. The underdamped equation is integrated with a second order Verlet-like integrator \cite{sivak2013using}, with a timestep of $h=10^{-3}$. Both nonequilibrium results and LDFs are calculated from trajectories of the length $t_N=8\times 10^5$. Fig. \ref{fig:tracer}$A$ and $B$ shows the LDF while fixing $\lambda_{Q_1}=0$ and $\lambda_J=0$, respectively. To evaluate $t_N\left \langle J^2 \right \rangle_0$, we compute the LDFs at $\lambda_J\in [-0.0035,0.0035]$, $\lambda_{Q_1}=0$ and fit the curve with a parabola to estimate its curvature. Similarly, $t_N\left \langle Q_1^2 \right \rangle_0$ is estimated with LDFs at $\lambda_J=0$, $\lambda_{Q_1}\in [-0.0035,0.0035]$. The statistical errors in LDFs are estimated by the standard deviation among 15 independent samples. To evaluate the statistical error in the linear transport coefficient, we fit a parabola to each of the sample sets, and compute the standard error of the mean among the 15 curvatures. The fitted curves yield an estimate of $t_N\left \langle J^2 \right \rangle_0/2=2.6874\pm0.0015$, $t_N\left \langle Q_1^2 \right \rangle_0/2=2.6773\pm0.0013$, and the mobility is $30.2\pm6.8$. 

In Fig. \ref{fig:tracer}$C$, we plot the theoretically predicted linear response along with the nonequilibrium simulation results to show their agreement at small values of $F$.  Nonequilibrium results are averaged over $2400$ independent trajectories and standard errors of the mean are plotted. This model has recently been shown to exhibit a negative differential mobility \cite{sarracino2016nonlinear}. Near equilibrium, the mobility is proportional to $\langle J^2 \rangle_0$, and thus must be non-negative. However, the linear response around a nonequilibrium steady state given in Eq.~\ref{Eq:LRSS2} clarifies how a negative differential mobility is possible. While not true at the conditions we consider, in principle $t_N\langle Q_1^2 \rangle_0 $ may be larger than $2/(\gamma\beta)+t_N\left \langle J^2 \right \rangle_0$, resulting in a current that decreases with added force.

\section*{Discussion and Conclusion}

We have introduced a method to compute arbitrarily high order transport coefficients that is heavily built on early works by Onsager and Kubo, and can be seen as an extension to stochastic systems in the modern language of stochastic thermodynamics. The relative action we introduce  acts as the generalized ``dissipation function'' as Onsager first invoked \cite{onsager1953fluctuations}, which gives the probability of a temporal succession of states. While in the linear response regime, the dissipation function is simply given by the rate of entropy production, to describe nonlinear response, additional information, namely the symmetric part of the action, is required. By constructing a LDF of both the time symmetric and asymmetric dynamical variables, we arrive at a function evaluated in equilibrium that contains all of the information about current-affinity relationship as encoded through microscopic correlations. 

A primary limitation of our method is the assumption that the nonequilibrium phase space distribution function can be expanded as a power series of the affinity. Critics of Kubo's nonlinear response theory have pointed out that for many transport processes, such expansions do not exist \cite{evans1983molecular,morriss2013statistical}. Indeed, a perturbative treatment breaks down when long-range or long-time correlations exist in a system in the thermodynamic limit. This can happen in the vicinity of phase transitions, where diverging correlation lengths will cause a divergence in the correlation functions corresponding to the transport coefficients. As a result, our simple series expansion is likely to fail in describing dynamical phase transitions between nonequilibrium steady states. Additionally, constraints on dynamics can also result in long-range correlations. For example, low dimensional molecular fluids are known to exhibit diverging diffusivity  computed from the Green-Kubo formula, which originates from the slowly decaying correlation of hydrodynamic modes associated with conserved quantities \cite{alder1970decay,dorfman1994generic}. In deterministic systems, conservation of energy and momentum confines the trajectories of the system to certain manifolds of the phase space. As a consequence, currents associated with conserved quantities are often non-analytic functions of the affinity \cite{kawasaki1973theory,yamada1975application}. To circumvent such problems, we have restricted ourselves to Markovian stochastic systems in the present work, where ergodicity is guaranteed. In all the examples shown, correlation functions decay to zero in microscopic timescales, so that their integrals in the long time limit are well defined. 

The method we propose is general enough that it can be applied to stochastic systems both near and far from equilibrium, regardless of the specific type of affinity, as long as a microscopic definition of current and path action can be written down. Unlike direct nonequilibrium simulation, our results do not depend on the details of the system, such as the nature of the noise, since correlation functions are evaluated in an equilibrium ensemble. Our method is expected to exhibit superior statistical performance compared to methods relying on transient time correlation functions, based on a detailed study on the statistical error in the computation of linear transport coefficients by the Green-Kubo formulas and our method \cite{gao2017transport}. Higher order transport coefficients demand a more accurate evaluation of the large deviation function. However, with the advancement of algorithms enhanced by importance sampling techniques, we expect that our method will become a standard approach to computing higher order transport coefficients. Furthermore, the expression of transport coefficients in terms of multi-time correlation functions builds a connection between macroscopic transport processes and microscopic observables. We expect that theoretical manipulations on these multi-time correlation functions will give us further insights into how nonlinear response behaviors arise from the molecular details of the system.

\section*{Appendix}
\subsection*{Derivation of the nonequilibrium average current for a general Langevin system}
Consider a system with $N$ degrees of freedom interacting through potential $V(\boldsymbol{r})$ at temperature $T$. Each degree of freedom $i$ evolves under the underdamped Langevin equation
\begin{equation}
m_i\dot{v}_i=-m_i\gamma v_i-\frac{dV(\boldsymbol{r})}{dr_i}+F+\eta_i,
\end{equation}
where $m_i$ is the mass, $\gamma$ is the friction coefficient, $F$ is a constant force, and $\eta_i$'s are Gaussian noise satisfying $\left \langle \eta_i(0)\eta_j(t)\right \rangle = 2m_i\gamma k_BT\delta(t)\delta_{ij}$. The probability of observing a given trajectory can be written in the Onsager-Machlup form \cite{onsager1953fluctuations}
\begin{equation}
\begin{aligned}
&P_F[\tilde{x}]\propto \\
&\exp \left [-\int dt\sum_{i=1}^N\frac{(m_i\dot{v}_i+m_i\gamma v_i+dV(\boldsymbol{r})/dr_i -F)^2}{4m_i\gamma k_BT} \right ]. \nonumber
\end{aligned}
\end{equation}

The relative stochastic action can be derived from the ratio of path probability with and without $F$,
\begin{equation}
U[\tilde{x}]=-\sum_{i=1}^N \frac{-(2m_i\dot{v}_i+2m_i\gamma v_i+2dV(\boldsymbol{r})/dr_i) F+F^2}{4m_i\gamma}.
\end{equation}
As only the velocities are asymmetric upon time reversal, we identify the time asymmetric and symmetric parts as
\begin{equation}
\begin{aligned}
A[\tilde{x}]&=\sum_{i=1}^N \frac{v_iF}{2}, \\
S[\tilde{x}]&=\sum_{i=1}^N\frac{(m_i\dot{v}_i+dV(\boldsymbol{r})/dr_i)F}{2m_i\gamma}-\sum_{i=1}^N\frac{F^2}{4m_i\gamma}.
\end{aligned}
\end{equation}
As seen, the time symmetric part has a part proportional to $F$, and a part proportional to $F^2$; in other words, $S[\Tilde{x}]=\sum_{i=1}^p t_NQ_iX^i$ where $p=2$ . Defining the affinity $X=F/2$, we arrive at the expressions of the dynamical variables,
\begin{equation}
j(\tilde{x}_t)=\sum_{i=1}^N v_i, \,q_1(\tilde{x}_t)=\sum_{i=1}^N\frac{m_i\dot{v}_i+dV(\boldsymbol{r})/dr_i}{m_i\gamma}.
\end{equation}
As before, $q_2(\tilde{x}_t)=-\sum_{i=1}^N1/{m_i\gamma}$ is a constant independent of the dynamics. Therefore, the expression for nonequilibrium current can be written in a closed form
\begin{equation}
\left \langle J \right \rangle_X=\sum_{n=1}^{\infty}\frac{(\beta t_N X)^{n-1}}{n!}\sum_{i+j=n} iC_n^i \kappa_{i,j}^n,
\end{equation}
where $\beta=1/k_BT$, and the second sum is taken over all combinations of nonnegative indices $i,j$ such that $i+j=n$. $\kappa_{i,j}^n$ denotes the $n$-th order cumulant, with the first few terms:
\begin{equation}
\begin{aligned}
&\kappa_{1,0}^1=\left\langle J\right\rangle_0, \kappa_{0,1}^1=\left\langle Q_1\right\rangle_0, \\
&\kappa_{2,0}^2=\left\langle(\delta J)^2\right\rangle_0, \kappa_{1,1}^2=\left\langle\delta J \delta Q_1 \right\rangle_0,\kappa_{0,2}^2=\left\langle(\delta Q_1)^2\right\rangle_0,\\
&\kappa_{3,0}^3=\left\langle(\delta J)^3\right\rangle_0, \kappa_{2,1}^3=\left\langle(\delta J)^2\delta Q_1 \right\rangle_0,\kappa_{1,2}^3=\left\langle \delta J(\delta Q_1 )^2\right\rangle_0,\\
&\kappa_{0,3}^3=\left\langle(\delta Q_1)^3\right\rangle_0\\
&\cdots
\end{aligned}
\end{equation}


\section*{Acknowledgements}
The authors thank the UC Berkeley College of Chemistry for support. CYG was supported by the Kavli Energy NanoScience Institute.

\section*{References}
\bibliography{sample.bib}

\begin{thebibliography}{61}%
\makeatletter
\providecommand \@ifxundefined [1]{%
 \@ifx{#1\undefined}
}%
\providecommand \@ifnum [1]{%
 \ifnum #1\expandafter \@firstoftwo
 \else \expandafter \@secondoftwo
 \fi
}%
\providecommand \@ifx [1]{%
 \ifx #1\expandafter \@firstoftwo
 \else \expandafter \@secondoftwo
 \fi
}%
\providecommand \natexlab [1]{#1}%
\providecommand \enquote  [1]{``#1''}%
\providecommand \bibnamefont  [1]{#1}%
\providecommand \bibfnamefont [1]{#1}%
\providecommand \citenamefont [1]{#1}%
\providecommand \href@noop [0]{\@secondoftwo}%
\providecommand \href [0]{\begingroup \@sanitize@url \@href}%
\providecommand \@href[1]{\@@startlink{#1}\@@href}%
\providecommand \@@href[1]{\endgroup#1\@@endlink}%
\providecommand \@sanitize@url [0]{\catcode `\\12\catcode `\$12\catcode
  `\&12\catcode `\#12\catcode `\^12\catcode `\_12\catcode `\%12\relax}%
\providecommand \@@startlink[1]{}%
\providecommand \@@endlink[0]{}%
\providecommand \url  [0]{\begingroup\@sanitize@url \@url }%
\providecommand \@url [1]{\endgroup\@href {#1}{\urlprefix }}%
\providecommand \urlprefix  [0]{URL }%
\providecommand \Eprint [0]{\href }%
\providecommand \doibase [0]{http://dx.doi.org/}%
\providecommand \selectlanguage [0]{\@gobble}%
\providecommand \bibinfo  [0]{\@secondoftwo}%
\providecommand \bibfield  [0]{\@secondoftwo}%
\providecommand \translation [1]{[#1]}%
\providecommand \BibitemOpen [0]{}%
\providecommand \bibitemStop [0]{}%
\providecommand \bibitemNoStop [0]{.\EOS\space}%
\providecommand \EOS [0]{\spacefactor3000\relax}%
\providecommand \BibitemShut  [1]{\csname bibitem#1\endcsname}%
\let\auto@bib@innerbib\@empty
\bibitem [{\citenamefont {Onsager}(1931)}]{onsager1931reciprocal}%
  \BibitemOpen
  \bibfield  {author} {\bibinfo {author} {\bibfnamefont {L.}~\bibnamefont
  {Onsager}},\ }\href@noop {} {\bibfield  {journal} {\bibinfo  {journal}
  {Physical review}\ }\textbf {\bibinfo {volume} {37}},\ \bibinfo {pages} {405}
  (\bibinfo {year} {1931})}\BibitemShut {NoStop}%
\bibitem [{\citenamefont {Onsager}\ and\ \citenamefont
  {Machlup}(1953)}]{onsager1953fluctuations}%
  \BibitemOpen
  \bibfield  {author} {\bibinfo {author} {\bibfnamefont {L.}~\bibnamefont
  {Onsager}}\ and\ \bibinfo {author} {\bibfnamefont {S.}~\bibnamefont
  {Machlup}},\ }\href@noop {} {\bibfield  {journal} {\bibinfo  {journal}
  {Physical Review}\ }\textbf {\bibinfo {volume} {91}},\ \bibinfo {pages}
  {1505} (\bibinfo {year} {1953})}\BibitemShut {NoStop}%
\bibitem [{\citenamefont {Kubo}(1957)}]{kubo1957statistical}%
  \BibitemOpen
  \bibfield  {author} {\bibinfo {author} {\bibfnamefont {R.}~\bibnamefont
  {Kubo}},\ }\href@noop {} {\bibfield  {journal} {\bibinfo  {journal} {Journal
  of the Physical Society of Japan}\ }\textbf {\bibinfo {volume} {12}},\
  \bibinfo {pages} {570} (\bibinfo {year} {1957})}\BibitemShut {NoStop}%
\bibitem [{\citenamefont {Green}(1954)}]{green1954markoff}%
  \BibitemOpen
  \bibfield  {author} {\bibinfo {author} {\bibfnamefont {M.~S.}\ \bibnamefont
  {Green}},\ }\href@noop {} {\bibfield  {journal} {\bibinfo  {journal} {The
  Journal of Chemical Physics}\ }\textbf {\bibinfo {volume} {22}},\ \bibinfo
  {pages} {398} (\bibinfo {year} {1954})}\BibitemShut {NoStop}%
\bibitem [{\citenamefont {Levesque}\ \emph {et~al.}(1973)\citenamefont
  {Levesque}, \citenamefont {Verlet},\ and\ \citenamefont
  {K{\"u}rkijarvi}}]{levesque1973computer}%
  \BibitemOpen
  \bibfield  {author} {\bibinfo {author} {\bibfnamefont {D.}~\bibnamefont
  {Levesque}}, \bibinfo {author} {\bibfnamefont {L.}~\bibnamefont {Verlet}}, \
  and\ \bibinfo {author} {\bibfnamefont {J.}~\bibnamefont {K{\"u}rkijarvi}},\
  }\href@noop {} {\bibfield  {journal} {\bibinfo  {journal} {Physical Review
  A}\ }\textbf {\bibinfo {volume} {7}},\ \bibinfo {pages} {1690} (\bibinfo
  {year} {1973})}\BibitemShut {NoStop}%
\bibitem [{\citenamefont {Schelling}\ \emph {et~al.}(2002)\citenamefont
  {Schelling}, \citenamefont {Phillpot},\ and\ \citenamefont
  {Keblinski}}]{schelling2002comparison}%
  \BibitemOpen
  \bibfield  {author} {\bibinfo {author} {\bibfnamefont {P.~K.}\ \bibnamefont
  {Schelling}}, \bibinfo {author} {\bibfnamefont {S.~R.}\ \bibnamefont
  {Phillpot}}, \ and\ \bibinfo {author} {\bibfnamefont {P.}~\bibnamefont
  {Keblinski}},\ }\href@noop {} {\bibfield  {journal} {\bibinfo  {journal}
  {Physical Review B}\ }\textbf {\bibinfo {volume} {65}},\ \bibinfo {pages}
  {144306} (\bibinfo {year} {2002})}\BibitemShut {NoStop}%
\bibitem [{\citenamefont {Tenenbaum}\ \emph {et~al.}(1982)\citenamefont
  {Tenenbaum}, \citenamefont {Ciccotti},\ and\ \citenamefont
  {Gallico}}]{tenenbaum1982stationary}%
  \BibitemOpen
  \bibfield  {author} {\bibinfo {author} {\bibfnamefont {A.}~\bibnamefont
  {Tenenbaum}}, \bibinfo {author} {\bibfnamefont {G.}~\bibnamefont {Ciccotti}},
  \ and\ \bibinfo {author} {\bibfnamefont {R.}~\bibnamefont {Gallico}},\
  }\href@noop {} {\bibfield  {journal} {\bibinfo  {journal} {Physical Review
  A}\ }\textbf {\bibinfo {volume} {25}},\ \bibinfo {pages} {2778} (\bibinfo
  {year} {1982})}\BibitemShut {NoStop}%
\bibitem [{\citenamefont {Baranyai}\ and\ \citenamefont
  {Cummings}(1999)}]{baranyai1999steady}%
  \BibitemOpen
  \bibfield  {author} {\bibinfo {author} {\bibfnamefont {A.}~\bibnamefont
  {Baranyai}}\ and\ \bibinfo {author} {\bibfnamefont {P.~T.}\ \bibnamefont
  {Cummings}},\ }\href@noop {} {\bibfield  {journal} {\bibinfo  {journal} {The
  Journal of chemical physics}\ }\textbf {\bibinfo {volume} {110}},\ \bibinfo
  {pages} {42} (\bibinfo {year} {1999})}\BibitemShut {NoStop}%
\bibitem [{\citenamefont {Hoover}\ \emph {et~al.}(1980)\citenamefont {Hoover},
  \citenamefont {Evans}, \citenamefont {Hickman}, \citenamefont {Ladd},
  \citenamefont {Ashurst},\ and\ \citenamefont {Moran}}]{hoover1980lennard}%
  \BibitemOpen
  \bibfield  {author} {\bibinfo {author} {\bibfnamefont {W.~G.}\ \bibnamefont
  {Hoover}}, \bibinfo {author} {\bibfnamefont {D.~J.}\ \bibnamefont {Evans}},
  \bibinfo {author} {\bibfnamefont {R.~B.}\ \bibnamefont {Hickman}}, \bibinfo
  {author} {\bibfnamefont {A.~J.}\ \bibnamefont {Ladd}}, \bibinfo {author}
  {\bibfnamefont {W.~T.}\ \bibnamefont {Ashurst}}, \ and\ \bibinfo {author}
  {\bibfnamefont {B.}~\bibnamefont {Moran}},\ }\href@noop {} {\bibfield
  {journal} {\bibinfo  {journal} {Physical Review A}\ }\textbf {\bibinfo
  {volume} {22}},\ \bibinfo {pages} {1690} (\bibinfo {year}
  {1980})}\BibitemShut {NoStop}%
\bibitem [{\citenamefont {Evans}(1982)}]{evans1982homogeneous}%
  \BibitemOpen
  \bibfield  {author} {\bibinfo {author} {\bibfnamefont {D.~J.}\ \bibnamefont
  {Evans}},\ }\href@noop {} {\bibfield  {journal} {\bibinfo  {journal} {Physics
  Letters A}\ }\textbf {\bibinfo {volume} {91}},\ \bibinfo {pages} {457}
  (\bibinfo {year} {1982})}\BibitemShut {NoStop}%
\bibitem [{\citenamefont {M{\"u}ller-Plathe}(1997)}]{muller1997simple}%
  \BibitemOpen
  \bibfield  {author} {\bibinfo {author} {\bibfnamefont {F.}~\bibnamefont
  {M{\"u}ller-Plathe}},\ }\href@noop {} {\bibfield  {journal} {\bibinfo
  {journal} {The Journal of chemical physics}\ }\textbf {\bibinfo {volume}
  {106}},\ \bibinfo {pages} {6082} (\bibinfo {year} {1997})}\BibitemShut
  {NoStop}%
\bibitem [{\citenamefont {Tuckerman}\ \emph {et~al.}(1997)\citenamefont
  {Tuckerman}, \citenamefont {Mundy}, \citenamefont {Balasubramanian},\ and\
  \citenamefont {Klein}}]{tuckerman1997modified}%
  \BibitemOpen
  \bibfield  {author} {\bibinfo {author} {\bibfnamefont {M.~E.}\ \bibnamefont
  {Tuckerman}}, \bibinfo {author} {\bibfnamefont {C.~J.}\ \bibnamefont
  {Mundy}}, \bibinfo {author} {\bibfnamefont {S.}~\bibnamefont
  {Balasubramanian}}, \ and\ \bibinfo {author} {\bibfnamefont {M.~L.}\
  \bibnamefont {Klein}},\ }\href@noop {} {\bibfield  {journal} {\bibinfo
  {journal} {The Journal of chemical physics}\ }\textbf {\bibinfo {volume}
  {106}},\ \bibinfo {pages} {5615} (\bibinfo {year} {1997})}\BibitemShut
  {NoStop}%
\bibitem [{\citenamefont {Tenney}\ and\ \citenamefont
  {Maginn}(2010)}]{tenney2010limitations}%
  \BibitemOpen
  \bibfield  {author} {\bibinfo {author} {\bibfnamefont {C.~M.}\ \bibnamefont
  {Tenney}}\ and\ \bibinfo {author} {\bibfnamefont {E.~J.}\ \bibnamefont
  {Maginn}},\ }\href@noop {} {\bibfield  {journal} {\bibinfo  {journal} {The
  Journal of chemical physics}\ }\textbf {\bibinfo {volume} {132}},\ \bibinfo
  {pages} {014103} (\bibinfo {year} {2010})}\BibitemShut {NoStop}%
\bibitem [{\citenamefont {Efremov}(1969)}]{efremov1969fluctuation}%
  \BibitemOpen
  \bibfield  {author} {\bibinfo {author} {\bibfnamefont {G.}~\bibnamefont
  {Efremov}},\ }\href@noop {} {\bibfield  {journal} {\bibinfo  {journal} {Sov.
  Phys. JETP}\ }\textbf {\bibinfo {volume} {28}},\ \bibinfo {pages} {1232}
  (\bibinfo {year} {1969})}\BibitemShut {NoStop}%
\bibitem [{\citenamefont {Stratonovich}(1970)}]{stratonovich1970contribution}%
  \BibitemOpen
  \bibfield  {author} {\bibinfo {author} {\bibfnamefont {R.}~\bibnamefont
  {Stratonovich}},\ }\href@noop {} {\bibfield  {journal} {\bibinfo  {journal}
  {SOVIET PHYSICS JETP}\ }\textbf {\bibinfo {volume} {31}} (\bibinfo {year}
  {1970})}\BibitemShut {NoStop}%
\bibitem [{\citenamefont {Yamada}\ and\ \citenamefont
  {Kawasaki}(1967)}]{yamada1967nonlinear}%
  \BibitemOpen
  \bibfield  {author} {\bibinfo {author} {\bibfnamefont {T.}~\bibnamefont
  {Yamada}}\ and\ \bibinfo {author} {\bibfnamefont {K.}~\bibnamefont
  {Kawasaki}},\ }\href@noop {} {\bibfield  {journal} {\bibinfo  {journal}
  {Progress of Theoretical Physics}\ }\textbf {\bibinfo {volume} {38}},\
  \bibinfo {pages} {1031} (\bibinfo {year} {1967})}\BibitemShut {NoStop}%
\bibitem [{\citenamefont {Bochkov}\ and\ \citenamefont
  {Kuzovlev}(1977)}]{bochkov1977general}%
  \BibitemOpen
  \bibfield  {author} {\bibinfo {author} {\bibfnamefont {G.}~\bibnamefont
  {Bochkov}}\ and\ \bibinfo {author} {\bibfnamefont {Y.~E.}\ \bibnamefont
  {Kuzovlev}},\ }\href@noop {} {\bibfield  {journal} {\bibinfo  {journal} {Zh.
  Eksp. Teor. Fiz}\ }\textbf {\bibinfo {volume} {72}},\ \bibinfo {pages} {238}
  (\bibinfo {year} {1977})}\BibitemShut {NoStop}%
\bibitem [{\citenamefont {Visscher}(1974)}]{visscher1974transport}%
  \BibitemOpen
  \bibfield  {author} {\bibinfo {author} {\bibfnamefont {W.~M.}\ \bibnamefont
  {Visscher}},\ }\href@noop {} {\bibfield  {journal} {\bibinfo  {journal}
  {Physical Review A}\ }\textbf {\bibinfo {volume} {10}},\ \bibinfo {pages}
  {2461} (\bibinfo {year} {1974})}\BibitemShut {NoStop}%
\bibitem [{\citenamefont {Dufty}\ and\ \citenamefont
  {Lindenfeld}(1979)}]{dufty1979nonlinear}%
  \BibitemOpen
  \bibfield  {author} {\bibinfo {author} {\bibfnamefont {J.~W.}\ \bibnamefont
  {Dufty}}\ and\ \bibinfo {author} {\bibfnamefont {M.~J.}\ \bibnamefont
  {Lindenfeld}},\ }\href@noop {} {\bibfield  {journal} {\bibinfo  {journal}
  {Journal of Statistical Physics}\ }\textbf {\bibinfo {volume} {20}},\
  \bibinfo {pages} {259} (\bibinfo {year} {1979})}\BibitemShut {NoStop}%
\bibitem [{\citenamefont {Cohen}(1983)}]{cohen1983kinetic}%
  \BibitemOpen
  \bibfield  {author} {\bibinfo {author} {\bibfnamefont {E.}~\bibnamefont
  {Cohen}},\ }\href@noop {} {\bibfield  {journal} {\bibinfo  {journal} {Physica
  A: Statistical Mechanics and its Applications}\ }\textbf {\bibinfo {volume}
  {118}},\ \bibinfo {pages} {17} (\bibinfo {year} {1983})}\BibitemShut
  {NoStop}%
\bibitem [{\citenamefont {Evans}\ and\ \citenamefont
  {Morriss}(1988)}]{evans1988transient}%
  \BibitemOpen
  \bibfield  {author} {\bibinfo {author} {\bibfnamefont {D.~J.}\ \bibnamefont
  {Evans}}\ and\ \bibinfo {author} {\bibfnamefont {G.~P.}\ \bibnamefont
  {Morriss}},\ }\href@noop {} {\bibfield  {journal} {\bibinfo  {journal}
  {Physical Review A}\ }\textbf {\bibinfo {volume} {38}},\ \bibinfo {pages}
  {4142} (\bibinfo {year} {1988})}\BibitemShut {NoStop}%
\bibitem [{\citenamefont {Jarzynski}(1997)}]{jarzynski1997nonequilibrium}%
  \BibitemOpen
  \bibfield  {author} {\bibinfo {author} {\bibfnamefont {C.}~\bibnamefont
  {Jarzynski}},\ }\href@noop {} {\bibfield  {journal} {\bibinfo  {journal}
  {Physical Review Letters}\ }\textbf {\bibinfo {volume} {78}},\ \bibinfo
  {pages} {2690} (\bibinfo {year} {1997})}\BibitemShut {NoStop}%
\bibitem [{\citenamefont {Crooks}(1999)}]{crooks1999entropy}%
  \BibitemOpen
  \bibfield  {author} {\bibinfo {author} {\bibfnamefont {G.~E.}\ \bibnamefont
  {Crooks}},\ }\href@noop {} {\bibfield  {journal} {\bibinfo  {journal}
  {Physical Review E}\ }\textbf {\bibinfo {volume} {60}},\ \bibinfo {pages}
  {2721} (\bibinfo {year} {1999})}\BibitemShut {NoStop}%
\bibitem [{\citenamefont {Gallavotti}\ and\ \citenamefont
  {Cohen}(1995)}]{gallavotti1995dynamical}%
  \BibitemOpen
  \bibfield  {author} {\bibinfo {author} {\bibfnamefont {G.}~\bibnamefont
  {Gallavotti}}\ and\ \bibinfo {author} {\bibfnamefont {E.~G.~D.}\ \bibnamefont
  {Cohen}},\ }\href@noop {} {\bibfield  {journal} {\bibinfo  {journal} {Journal
  of Statistical Physics}\ }\textbf {\bibinfo {volume} {80}},\ \bibinfo {pages}
  {931} (\bibinfo {year} {1995})}\BibitemShut {NoStop}%
\bibitem [{\citenamefont {Barato}\ and\ \citenamefont
  {Seifert}(2015)}]{barato2015thermodynamic}%
  \BibitemOpen
  \bibfield  {author} {\bibinfo {author} {\bibfnamefont {A.~C.}\ \bibnamefont
  {Barato}}\ and\ \bibinfo {author} {\bibfnamefont {U.}~\bibnamefont
  {Seifert}},\ }\href@noop {} {\bibfield  {journal} {\bibinfo  {journal}
  {Physical review letters}\ }\textbf {\bibinfo {volume} {114}},\ \bibinfo
  {pages} {158101} (\bibinfo {year} {2015})}\BibitemShut {NoStop}%
\bibitem [{\citenamefont {Gingrich}\ \emph {et~al.}(2016)\citenamefont
  {Gingrich}, \citenamefont {Horowitz}, \citenamefont {Perunov},\ and\
  \citenamefont {England}}]{gingrich2016dissipation}%
  \BibitemOpen
  \bibfield  {author} {\bibinfo {author} {\bibfnamefont {T.~R.}\ \bibnamefont
  {Gingrich}}, \bibinfo {author} {\bibfnamefont {J.~M.}\ \bibnamefont
  {Horowitz}}, \bibinfo {author} {\bibfnamefont {N.}~\bibnamefont {Perunov}}, \
  and\ \bibinfo {author} {\bibfnamefont {J.~L.}\ \bibnamefont {England}},\
  }\href@noop {} {\bibfield  {journal} {\bibinfo  {journal} {Physical review
  letters}\ }\textbf {\bibinfo {volume} {116}},\ \bibinfo {pages} {120601}
  (\bibinfo {year} {2016})}\BibitemShut {NoStop}%
\bibitem [{\citenamefont {Harada}\ and\ \citenamefont
  {Sasa}(2005)}]{harada2005equality}%
  \BibitemOpen
  \bibfield  {author} {\bibinfo {author} {\bibfnamefont {T.}~\bibnamefont
  {Harada}}\ and\ \bibinfo {author} {\bibfnamefont {S.-i.}\ \bibnamefont
  {Sasa}},\ }\href@noop {} {\bibfield  {journal} {\bibinfo  {journal} {Physical
  review letters}\ }\textbf {\bibinfo {volume} {95}},\ \bibinfo {pages}
  {130602} (\bibinfo {year} {2005})}\BibitemShut {NoStop}%
\bibitem [{\citenamefont {Speck}\ and\ \citenamefont
  {Seifert}(2006)}]{speck2006restoring}%
  \BibitemOpen
  \bibfield  {author} {\bibinfo {author} {\bibfnamefont {T.}~\bibnamefont
  {Speck}}\ and\ \bibinfo {author} {\bibfnamefont {U.}~\bibnamefont
  {Seifert}},\ }\href@noop {} {\bibfield  {journal} {\bibinfo  {journal} {EPL
  (Europhysics Letters)}\ }\textbf {\bibinfo {volume} {74}},\ \bibinfo {pages}
  {391} (\bibinfo {year} {2006})}\BibitemShut {NoStop}%
\bibitem [{\citenamefont {Speck}\ and\ \citenamefont
  {Seifert}(2009)}]{speck2009extended}%
  \BibitemOpen
  \bibfield  {author} {\bibinfo {author} {\bibfnamefont {T.}~\bibnamefont
  {Speck}}\ and\ \bibinfo {author} {\bibfnamefont {U.}~\bibnamefont
  {Seifert}},\ }\href@noop {} {\bibfield  {journal} {\bibinfo  {journal}
  {Physical Review E}\ }\textbf {\bibinfo {volume} {79}},\ \bibinfo {pages}
  {040102} (\bibinfo {year} {2009})}\BibitemShut {NoStop}%
\bibitem [{\citenamefont {Baiesi}\ \emph {et~al.}(2009)\citenamefont {Baiesi},
  \citenamefont {Maes},\ and\ \citenamefont
  {Wynants}}]{baiesi2009fluctuations}%
  \BibitemOpen
  \bibfield  {author} {\bibinfo {author} {\bibfnamefont {M.}~\bibnamefont
  {Baiesi}}, \bibinfo {author} {\bibfnamefont {C.}~\bibnamefont {Maes}}, \ and\
  \bibinfo {author} {\bibfnamefont {B.}~\bibnamefont {Wynants}},\ }\href@noop
  {} {\bibfield  {journal} {\bibinfo  {journal} {Physical review letters}\
  }\textbf {\bibinfo {volume} {103}},\ \bibinfo {pages} {010602} (\bibinfo
  {year} {2009})}\BibitemShut {NoStop}%
\bibitem [{\citenamefont {Basu}\ and\ \citenamefont
  {Maes}(2015)}]{basu2015nonequilibrium}%
  \BibitemOpen
  \bibfield  {author} {\bibinfo {author} {\bibfnamefont {U.}~\bibnamefont
  {Basu}}\ and\ \bibinfo {author} {\bibfnamefont {C.}~\bibnamefont {Maes}},\
  }\href@noop {} {\bibfield  {journal} {\bibinfo  {journal} {Journal of
  Physics}\ }\textbf {\bibinfo {volume} {638}},\ \bibinfo {pages} {012001}
  (\bibinfo {year} {2015})}\BibitemShut {NoStop}%
\bibitem [{\citenamefont {Andrieux}\ and\ \citenamefont
  {Gaspard}(2004)}]{andrieux2004fluctuation}%
  \BibitemOpen
  \bibfield  {author} {\bibinfo {author} {\bibfnamefont {D.}~\bibnamefont
  {Andrieux}}\ and\ \bibinfo {author} {\bibfnamefont {P.}~\bibnamefont
  {Gaspard}},\ }\href@noop {} {\bibfield  {journal} {\bibinfo  {journal} {The
  Journal of chemical physics}\ }\textbf {\bibinfo {volume} {121}},\ \bibinfo
  {pages} {6167} (\bibinfo {year} {2004})}\BibitemShut {NoStop}%
\bibitem [{\citenamefont {Gaspard}(2013)}]{gaspard2013multivariate}%
  \BibitemOpen
  \bibfield  {author} {\bibinfo {author} {\bibfnamefont {P.}~\bibnamefont
  {Gaspard}},\ }\href@noop {} {\bibfield  {journal} {\bibinfo  {journal} {New
  Journal of Physics}\ }\textbf {\bibinfo {volume} {15}},\ \bibinfo {pages}
  {115014} (\bibinfo {year} {2013})}\BibitemShut {NoStop}%
\bibitem [{\citenamefont {Ellis}(2007)}]{ellis2007entropy}%
  \BibitemOpen
  \bibfield  {author} {\bibinfo {author} {\bibfnamefont {R.~S.}\ \bibnamefont
  {Ellis}},\ }\href@noop {} {\emph {\bibinfo {title} {Entropy, large
  deviations, and statistical mechanics}}}\ (\bibinfo  {publisher} {Springer},\
  \bibinfo {year} {2007})\BibitemShut {NoStop}%
\bibitem [{\citenamefont {Touchette}(2009)}]{touchette2009large}%
  \BibitemOpen
  \bibfield  {author} {\bibinfo {author} {\bibfnamefont {H.}~\bibnamefont
  {Touchette}},\ }\href@noop {} {\bibfield  {journal} {\bibinfo  {journal}
  {Physics Reports}\ }\textbf {\bibinfo {volume} {478}},\ \bibinfo {pages} {1}
  (\bibinfo {year} {2009})}\BibitemShut {NoStop}%
\bibitem [{\citenamefont {Hedges}\ \emph {et~al.}(2009)\citenamefont {Hedges},
  \citenamefont {Jack}, \citenamefont {Garrahan},\ and\ \citenamefont
  {Chandler}}]{hedges2009dynamic}%
  \BibitemOpen
  \bibfield  {author} {\bibinfo {author} {\bibfnamefont {L.~O.}\ \bibnamefont
  {Hedges}}, \bibinfo {author} {\bibfnamefont {R.~L.}\ \bibnamefont {Jack}},
  \bibinfo {author} {\bibfnamefont {J.~P.}\ \bibnamefont {Garrahan}}, \ and\
  \bibinfo {author} {\bibfnamefont {D.}~\bibnamefont {Chandler}},\ }\href@noop
  {} {\bibfield  {journal} {\bibinfo  {journal} {Science}\ }\textbf {\bibinfo
  {volume} {323}},\ \bibinfo {pages} {1309} (\bibinfo {year}
  {2009})}\BibitemShut {NoStop}%
\bibitem [{\citenamefont {Hurtado}\ and\ \citenamefont
  {Garrido}(2011)}]{hurtado2011spontaneous}%
  \BibitemOpen
  \bibfield  {author} {\bibinfo {author} {\bibfnamefont {P.~I.}\ \bibnamefont
  {Hurtado}}\ and\ \bibinfo {author} {\bibfnamefont {P.~L.}\ \bibnamefont
  {Garrido}},\ }\href@noop {} {\bibfield  {journal} {\bibinfo  {journal}
  {Physical review letters}\ }\textbf {\bibinfo {volume} {107}},\ \bibinfo
  {pages} {180601} (\bibinfo {year} {2011})}\BibitemShut {NoStop}%
\bibitem [{\citenamefont {Limmer}\ and\ \citenamefont
  {Chandler}(2014)}]{limmer2014theory}%
  \BibitemOpen
  \bibfield  {author} {\bibinfo {author} {\bibfnamefont {D.~T.}\ \bibnamefont
  {Limmer}}\ and\ \bibinfo {author} {\bibfnamefont {D.}~\bibnamefont
  {Chandler}},\ }\href@noop {} {\bibfield  {journal} {\bibinfo  {journal}
  {Proceedings of the National Academy of Sciences}\ ,\ \bibinfo {pages}
  {201407277}} (\bibinfo {year} {2014})}\BibitemShut {NoStop}%
\bibitem [{\citenamefont {Pre}\ and\ \citenamefont
  {Limmer}(2018)}]{pre2018current}%
  \BibitemOpen
  \bibfield  {author} {\bibinfo {author} {\bibfnamefont {T.~G.}\ \bibnamefont
  {Pre}}\ and\ \bibinfo {author} {\bibfnamefont {D.~T.}\ \bibnamefont
  {Limmer}},\ }\href@noop {} {\bibfield  {journal} {\bibinfo  {journal} {arXiv
  preprint arXiv:1805.04122}\ } (\bibinfo {year} {2018})}\BibitemShut {NoStop}%
\bibitem [{\citenamefont {Gao}\ and\ \citenamefont
  {Limmer}(2017)}]{gao2017transport}%
  \BibitemOpen
  \bibfield  {author} {\bibinfo {author} {\bibfnamefont {C.~Y.}\ \bibnamefont
  {Gao}}\ and\ \bibinfo {author} {\bibfnamefont {D.~T.}\ \bibnamefont
  {Limmer}},\ }\href@noop {} {\bibfield  {journal} {\bibinfo  {journal}
  {Entropy}\ }\textbf {\bibinfo {volume} {19}},\ \bibinfo {pages} {571}
  (\bibinfo {year} {2017})}\BibitemShut {NoStop}%
\bibitem [{\citenamefont {Maes}\ \emph {et~al.}(2008)\citenamefont {Maes},
  \citenamefont {Neto{\v{c}}n{\`y}},\ and\ \citenamefont
  {Wynants}}]{maes2008steady}%
  \BibitemOpen
  \bibfield  {author} {\bibinfo {author} {\bibfnamefont {C.}~\bibnamefont
  {Maes}}, \bibinfo {author} {\bibfnamefont {K.}~\bibnamefont
  {Neto{\v{c}}n{\`y}}}, \ and\ \bibinfo {author} {\bibfnamefont
  {B.}~\bibnamefont {Wynants}},\ }\href@noop {} {\bibfield  {journal} {\bibinfo
   {journal} {Physica A: Statistical Mechanics and its Applications}\ }\textbf
  {\bibinfo {volume} {387}},\ \bibinfo {pages} {2675} (\bibinfo {year}
  {2008})}\BibitemShut {NoStop}%
\bibitem [{\citenamefont {Nemoto}\ and\ \citenamefont
  {Sasa}(2014)}]{nemoto2014computation}%
  \BibitemOpen
  \bibfield  {author} {\bibinfo {author} {\bibfnamefont {T.}~\bibnamefont
  {Nemoto}}\ and\ \bibinfo {author} {\bibfnamefont {S.-i.}\ \bibnamefont
  {Sasa}},\ }\href@noop {} {\bibfield  {journal} {\bibinfo  {journal} {Physical
  review letters}\ }\textbf {\bibinfo {volume} {112}},\ \bibinfo {pages}
  {090602} (\bibinfo {year} {2014})}\BibitemShut {NoStop}%
\bibitem [{\citenamefont {Nemoto}\ \emph {et~al.}(2017)\citenamefont {Nemoto},
  \citenamefont {Jack},\ and\ \citenamefont {Lecomte}}]{nemoto2017finite}%
  \BibitemOpen
  \bibfield  {author} {\bibinfo {author} {\bibfnamefont {T.}~\bibnamefont
  {Nemoto}}, \bibinfo {author} {\bibfnamefont {R.~L.}\ \bibnamefont {Jack}}, \
  and\ \bibinfo {author} {\bibfnamefont {V.}~\bibnamefont {Lecomte}},\
  }\href@noop {} {\bibfield  {journal} {\bibinfo  {journal} {Physical review
  letters}\ }\textbf {\bibinfo {volume} {118}},\ \bibinfo {pages} {115702}
  (\bibinfo {year} {2017})}\BibitemShut {NoStop}%
\bibitem [{\citenamefont {Klymko}\ \emph {et~al.}(2018)\citenamefont {Klymko},
  \citenamefont {Geissler}, \citenamefont {Garrahan},\ and\ \citenamefont
  {Whitelam}}]{klymko2018rare}%
  \BibitemOpen
  \bibfield  {author} {\bibinfo {author} {\bibfnamefont {K.}~\bibnamefont
  {Klymko}}, \bibinfo {author} {\bibfnamefont {P.~L.}\ \bibnamefont
  {Geissler}}, \bibinfo {author} {\bibfnamefont {J.~P.}\ \bibnamefont
  {Garrahan}}, \ and\ \bibinfo {author} {\bibfnamefont {S.}~\bibnamefont
  {Whitelam}},\ }\href@noop {} {\bibfield  {journal} {\bibinfo  {journal}
  {Physical Review E}\ }\textbf {\bibinfo {volume} {97}},\ \bibinfo {pages}
  {032123} (\bibinfo {year} {2018})}\BibitemShut {NoStop}%
\bibitem [{\citenamefont {Ray}\ \emph {et~al.}(2018)\citenamefont {Ray},
  \citenamefont {Chan},\ and\ \citenamefont {Limmer}}]{ray2018exact}%
  \BibitemOpen
  \bibfield  {author} {\bibinfo {author} {\bibfnamefont {U.}~\bibnamefont
  {Ray}}, \bibinfo {author} {\bibfnamefont {G.~K.-L.}\ \bibnamefont {Chan}}, \
  and\ \bibinfo {author} {\bibfnamefont {D.~T.}\ \bibnamefont {Limmer}},\
  }\href@noop {} {\bibfield  {journal} {\bibinfo  {journal} {Physical review
  letters}\ }\textbf {\bibinfo {volume} {120}},\ \bibinfo {pages} {210602}
  (\bibinfo {year} {2018})}\BibitemShut {NoStop}%
\bibitem [{\citenamefont {Colangeli}\ \emph {et~al.}(2011)\citenamefont
  {Colangeli}, \citenamefont {Maes},\ and\ \citenamefont
  {Wynants}}]{colangeli2011meaningful}%
  \BibitemOpen
  \bibfield  {author} {\bibinfo {author} {\bibfnamefont {M.}~\bibnamefont
  {Colangeli}}, \bibinfo {author} {\bibfnamefont {C.}~\bibnamefont {Maes}}, \
  and\ \bibinfo {author} {\bibfnamefont {B.}~\bibnamefont {Wynants}},\
  }\href@noop {} {\bibfield  {journal} {\bibinfo  {journal} {Journal of Physics
  A: Mathematical and Theoretical}\ }\textbf {\bibinfo {volume} {44}},\
  \bibinfo {pages} {095001} (\bibinfo {year} {2011})}\BibitemShut {NoStop}%
\bibitem [{\citenamefont {Basu}\ \emph {et~al.}(2015)\citenamefont {Basu},
  \citenamefont {Kr{\"u}ger}, \citenamefont {Lazarescu},\ and\ \citenamefont
  {Maes}}]{basu2015frenetic}%
  \BibitemOpen
  \bibfield  {author} {\bibinfo {author} {\bibfnamefont {U.}~\bibnamefont
  {Basu}}, \bibinfo {author} {\bibfnamefont {M.}~\bibnamefont {Kr{\"u}ger}},
  \bibinfo {author} {\bibfnamefont {A.}~\bibnamefont {Lazarescu}}, \ and\
  \bibinfo {author} {\bibfnamefont {C.}~\bibnamefont {Maes}},\ }\href@noop {}
  {\bibfield  {journal} {\bibinfo  {journal} {Physical Chemistry Chemical
  Physics}\ }\textbf {\bibinfo {volume} {17}},\ \bibinfo {pages} {6653}
  (\bibinfo {year} {2015})}\BibitemShut {NoStop}%
\bibitem [{\citenamefont {Freed}(1968)}]{freed1968generalized}%
  \BibitemOpen
  \bibfield  {author} {\bibinfo {author} {\bibfnamefont {J.~H.}\ \bibnamefont
  {Freed}},\ }\href@noop {} {\bibfield  {journal} {\bibinfo  {journal} {The
  Journal of Chemical Physics}\ }\textbf {\bibinfo {volume} {49}},\ \bibinfo
  {pages} {376} (\bibinfo {year} {1968})}\BibitemShut {NoStop}%
\bibitem [{\citenamefont {Wang}\ \emph {et~al.}(1998)\citenamefont {Wang},
  \citenamefont {Stettler}, \citenamefont {Yu},\ and\ \citenamefont
  {Maziar}}]{wang1998application}%
  \BibitemOpen
  \bibfield  {author} {\bibinfo {author} {\bibfnamefont {E.~X.}\ \bibnamefont
  {Wang}}, \bibinfo {author} {\bibfnamefont {M.}~\bibnamefont {Stettler}},
  \bibinfo {author} {\bibfnamefont {S.}~\bibnamefont {Yu}}, \ and\ \bibinfo
  {author} {\bibfnamefont {C.}~\bibnamefont {Maziar}},\ }in\ \href@noop {}
  {\emph {\bibinfo {booktitle} {1998 Sixth International Workshop on
  Computational Electronics. Extended Abstracts (Cat. No. 98EX116)}}}\
  (\bibinfo {organization} {IEEE},\ \bibinfo {year} {1998})\ pp.\ \bibinfo
  {pages} {234--237}\BibitemShut {NoStop}%
\bibitem [{\citenamefont {Giardina}\ \emph {et~al.}(2006)\citenamefont
  {Giardina}, \citenamefont {Kurchan},\ and\ \citenamefont
  {Peliti}}]{giardina2006direct}%
  \BibitemOpen
  \bibfield  {author} {\bibinfo {author} {\bibfnamefont {C.}~\bibnamefont
  {Giardina}}, \bibinfo {author} {\bibfnamefont {J.}~\bibnamefont {Kurchan}}, \
  and\ \bibinfo {author} {\bibfnamefont {L.}~\bibnamefont {Peliti}},\
  }\href@noop {} {\bibfield  {journal} {\bibinfo  {journal} {Physical review
  letters}\ }\textbf {\bibinfo {volume} {96}},\ \bibinfo {pages} {120603}
  (\bibinfo {year} {2006})}\BibitemShut {NoStop}%
\bibitem [{\citenamefont {Honeycutt}(1992)}]{honeycutt1992stochastic}%
  \BibitemOpen
  \bibfield  {author} {\bibinfo {author} {\bibfnamefont {R.~L.}\ \bibnamefont
  {Honeycutt}},\ }\href@noop {} {\bibfield  {journal} {\bibinfo  {journal}
  {Physical Review A}\ }\textbf {\bibinfo {volume} {45}},\ \bibinfo {pages}
  {600} (\bibinfo {year} {1992})}\BibitemShut {NoStop}%
\bibitem [{\citenamefont {Yang}\ \emph {et~al.}(2007)\citenamefont {Yang},
  \citenamefont {Li}, \citenamefont {Wang},\ and\ \citenamefont
  {Li}}]{yang2007thermal}%
  \BibitemOpen
  \bibfield  {author} {\bibinfo {author} {\bibfnamefont {N.}~\bibnamefont
  {Yang}}, \bibinfo {author} {\bibfnamefont {N.}~\bibnamefont {Li}}, \bibinfo
  {author} {\bibfnamefont {L.}~\bibnamefont {Wang}}, \ and\ \bibinfo {author}
  {\bibfnamefont {B.}~\bibnamefont {Li}},\ }\href@noop {} {\bibfield  {journal}
  {\bibinfo  {journal} {Physical Review B}\ }\textbf {\bibinfo {volume} {76}},\
  \bibinfo {pages} {020301} (\bibinfo {year} {2007})}\BibitemShut {NoStop}%
\bibitem [{\citenamefont {Frenkel}\ and\ \citenamefont
  {Smit}(2001)}]{frenkel2001understanding}%
  \BibitemOpen
  \bibfield  {author} {\bibinfo {author} {\bibfnamefont {D.}~\bibnamefont
  {Frenkel}}\ and\ \bibinfo {author} {\bibfnamefont {B.}~\bibnamefont {Smit}},\
  }\href@noop {} {\emph {\bibinfo {title} {Understanding molecular simulation:
  from algorithms to applications}}},\ Vol.~\bibinfo {volume} {1}\ (\bibinfo
  {publisher} {Elsevier},\ \bibinfo {year} {2001})\BibitemShut {NoStop}%
\bibitem [{\citenamefont {Sivak}\ \emph {et~al.}(2013)\citenamefont {Sivak},
  \citenamefont {Chodera},\ and\ \citenamefont {Crooks}}]{sivak2013using}%
  \BibitemOpen
  \bibfield  {author} {\bibinfo {author} {\bibfnamefont {D.~A.}\ \bibnamefont
  {Sivak}}, \bibinfo {author} {\bibfnamefont {J.~D.}\ \bibnamefont {Chodera}},
  \ and\ \bibinfo {author} {\bibfnamefont {G.~E.}\ \bibnamefont {Crooks}},\
  }\href@noop {} {\bibfield  {journal} {\bibinfo  {journal} {Physical Review
  X}\ }\textbf {\bibinfo {volume} {3}},\ \bibinfo {pages} {011007} (\bibinfo
  {year} {2013})}\BibitemShut {NoStop}%
\bibitem [{\citenamefont {Sarracino}\ \emph {et~al.}(2016)\citenamefont
  {Sarracino}, \citenamefont {Cecconi}, \citenamefont {Puglisi},\ and\
  \citenamefont {Vulpiani}}]{sarracino2016nonlinear}%
  \BibitemOpen
  \bibfield  {author} {\bibinfo {author} {\bibfnamefont {A.}~\bibnamefont
  {Sarracino}}, \bibinfo {author} {\bibfnamefont {F.}~\bibnamefont {Cecconi}},
  \bibinfo {author} {\bibfnamefont {A.}~\bibnamefont {Puglisi}}, \ and\
  \bibinfo {author} {\bibfnamefont {A.}~\bibnamefont {Vulpiani}},\ }\href@noop
  {} {\bibfield  {journal} {\bibinfo  {journal} {Physical review letters}\
  }\textbf {\bibinfo {volume} {117}},\ \bibinfo {pages} {174501} (\bibinfo
  {year} {2016})}\BibitemShut {NoStop}%
\bibitem [{\citenamefont {Evans}(1983)}]{evans1983molecular}%
  \BibitemOpen
  \bibfield  {author} {\bibinfo {author} {\bibfnamefont {D.~J.}\ \bibnamefont
  {Evans}},\ }\href@noop {} {\bibfield  {journal} {\bibinfo  {journal} {Physica
  A: Statistical Mechanics and its Applications}\ }\textbf {\bibinfo {volume}
  {118}},\ \bibinfo {pages} {51} (\bibinfo {year} {1983})}\BibitemShut
  {NoStop}%
\bibitem [{\citenamefont {Morriss}\ and\ \citenamefont
  {Evans}(2013)}]{morriss2013statistical}%
  \BibitemOpen
  \bibfield  {author} {\bibinfo {author} {\bibfnamefont {G.~P.}\ \bibnamefont
  {Morriss}}\ and\ \bibinfo {author} {\bibfnamefont {D.~J.}\ \bibnamefont
  {Evans}},\ }\href@noop {} {\emph {\bibinfo {title} {Statistical Mechanics of
  Nonequilbrium Liquids}}}\ (\bibinfo  {publisher} {ANU Press},\ \bibinfo
  {year} {2013})\BibitemShut {NoStop}%
\bibitem [{\citenamefont {Alder}\ and\ \citenamefont
  {Wainwright}(1970)}]{alder1970decay}%
  \BibitemOpen
  \bibfield  {author} {\bibinfo {author} {\bibfnamefont {B.}~\bibnamefont
  {Alder}}\ and\ \bibinfo {author} {\bibfnamefont {T.}~\bibnamefont
  {Wainwright}},\ }\href@noop {} {\bibfield  {journal} {\bibinfo  {journal}
  {Physical review A}\ }\textbf {\bibinfo {volume} {1}},\ \bibinfo {pages} {18}
  (\bibinfo {year} {1970})}\BibitemShut {NoStop}%
\bibitem [{\citenamefont {Dorfman}\ \emph {et~al.}(1994)\citenamefont
  {Dorfman}, \citenamefont {Kirkpatrick},\ and\ \citenamefont
  {Sengers}}]{dorfman1994generic}%
  \BibitemOpen
  \bibfield  {author} {\bibinfo {author} {\bibfnamefont {J.}~\bibnamefont
  {Dorfman}}, \bibinfo {author} {\bibfnamefont {T.}~\bibnamefont
  {Kirkpatrick}}, \ and\ \bibinfo {author} {\bibfnamefont {J.}~\bibnamefont
  {Sengers}},\ }\href@noop {} {\bibfield  {journal} {\bibinfo  {journal}
  {Annual Review of Physical Chemistry}\ }\textbf {\bibinfo {volume} {45}},\
  \bibinfo {pages} {213} (\bibinfo {year} {1994})}\BibitemShut {NoStop}%
\bibitem [{\citenamefont {Kawasaki}\ and\ \citenamefont
  {Gunton}(1973)}]{kawasaki1973theory}%
  \BibitemOpen
  \bibfield  {author} {\bibinfo {author} {\bibfnamefont {K.}~\bibnamefont
  {Kawasaki}}\ and\ \bibinfo {author} {\bibfnamefont {J.~D.}\ \bibnamefont
  {Gunton}},\ }\href@noop {} {\bibfield  {journal} {\bibinfo  {journal}
  {Physical Review A}\ }\textbf {\bibinfo {volume} {8}},\ \bibinfo {pages}
  {2048} (\bibinfo {year} {1973})}\BibitemShut {NoStop}%
\bibitem [{\citenamefont {Yamada}\ and\ \citenamefont
  {Kawasaki}(1975)}]{yamada1975application}%
  \BibitemOpen
  \bibfield  {author} {\bibinfo {author} {\bibfnamefont {T.}~\bibnamefont
  {Yamada}}\ and\ \bibinfo {author} {\bibfnamefont {K.}~\bibnamefont
  {Kawasaki}},\ }\href@noop {} {\bibfield  {journal} {\bibinfo  {journal}
  {Progress of theoretical physics}\ }\textbf {\bibinfo {volume} {53}},\
  \bibinfo {pages} {111} (\bibinfo {year} {1975})}\BibitemShut {NoStop}%
\end{thebibliography}%

\end{document}